\documentclass[12pt,journal,onecolumn,draftclsnofoot]{IEEEtran} 
\usepackage{graphicx}  
\usepackage{amsmath}   
\usepackage{amsthm}
\usepackage{bm}
\usepackage{amsfonts}
\usepackage{algorithmic}
\usepackage[ruled,noend]{algorithm2e}
\usepackage{epstopdf} 
\usepackage{url}
\usepackage{subcaption}
\usepackage{xcolor}
\usepackage{cite} 
\usepackage{tablefootnote}
\renewcommand\P{\mathbb P}

\newtheorem{prop} {Proposition}

\theoremstyle{definition}
\newtheorem{defn}{Problem Addressed}%[section]
\theoremstyle{remark}
\newtheorem{rem}{Remark}

\title{AMON: An Open Source Architecture for Online Monitoring, Statistical Analysis and Forensics of Multi-gigabit Streams}

\author{
 \IEEEauthorblockN{
  Michael Kallitsis,
  Stilian Stoev,
  Shrijita Bhattacharya,
  George Michailidis\
 }
 \IEEEauthorblockA{mgkallit@merit.edu, sstoev@umich.edu, shrijita@umich.edu, gmichail@ufl.edu}
}

\begin{document}

\maketitle
\vspace{-50pt}
\begin{abstract}
\vspace{-15pt}
The Internet, as a global system of interconnected networks, carries an extensive array of
information resources and services. Key requirements include good quality-of-service  and protection of 
the infrastructure from nefarious activity
 (e.g.\ distributed denial of service---DDoS---attacks).
 Network monitoring 
is essential to network engineering, capacity planning and prevention / mitigation of threats. We develop
an open source architecture, AMON (All-packet MONitor), for online monitoring and analysis of multi-gigabit network streams.  It leverages 
the high-performance packet monitor PF\_RING  and is readily deployable on commodity hardware. 
AMON examines {\em all packets}, partitions traffic into sub-streams by using rapid 
hashing and computes certain real-time data products. The resulting data structures provide 
views of the intensity {\em and} connectivity structure of network traffic  {\em at the time-scale of routing}.
The proposed \emph{integrated} framework includes modules for the \emph{identification}
of heavy-hitters as well as for \emph{visualization} and statistical \emph{detection}
 {\em at the time-of-onset}  of high impact events such as DDoS. 
This allows operators to quickly visualize and diagnose attacks, and limit offline and time-consuming post-mortem analysis.  
We demonstrate our system in the context of real-world attack incidents, and validate it against
state-of-the-art alternatives. 
AMON  has been deployed and is currently processing 10Gbps+ live Internet traffic at Merit Network. It is extensible 
and allows the addition of further statistical and filtering modules for real-time forensics.
 \end{abstract}
\vspace{-10pt}
\begin{IEEEkeywords}
\vspace{-10pt}
Network monitoring, detection, identification, visualization, PF\_RING, gigabit streams, commodity hardware, data products, 
algorithms, statistics, heavy tails, extreme value distribution, network attacks.
\end{IEEEkeywords}
\vspace{-25pt}
\section{Introduction}
%\subsection{Motivation and background}
\noindent\textbf{Motivation and background:} 
The Internet has become a vital resource to business, governments and society, worldwide. 
It has thrived and grown under diverse conditions and technologies with little to no centralized regulation.  
Its fundamental design principles 
have successfully ensured its robustness and broad accessibility. However, these same principles 
do not provide centralized management and/or monitoring of the {\em entire network}.  Therefore,
understanding broad based patterns such as traffic loads and thus adequacy of capacity and quality-of-service,
 composition of network traffic, adoption of new protocols and applications, are challenging tasks.
Such traffic  characterization problems and the corresponding network engineering, management and capacity planning solutions
made necessary the analysis of large volumes of data, and also gave rise to statistical techniques to handle them, such
as streaming algorithms \cite{gilbert:kotidis:muthukrishnan:strauss:2001,Muthukrishnan:2005:DSA:1166409.1166410}, sketches  
\cite{Krishnamurthy:2003:SCD:948205.948236,Gilbert:2007:OSF:1250790.1250824,4221749}, tomography \cite{xi:michailidis:nair:2006,lawrence:michailidis:nair:2006} and analysis of  heavy tails and long range dependence \cite{stoev:taqqu:park:michailidis:marron:2006,stoev:michailidis:2010}. %stoev:taqqu:park:marron:2005

In addition, it enables numerous vulnerabilities and security threats,  both to the infrastructure and to its user base.
For example, malicious activities such as {\em distributed denial of service} (DDoS) attacks are relatively 
easy to implement and rather hard to prevent,
since best practices like origin IP anti-spoofing (e.g., BCP38 recommendation~\cite{BCP38})
are not universally deployed by network operators. Their timely detection at appropriate short time-scales (e.g. in seconds)
 requires  processing vast amounts  of meta-data (e.g., NetFlow) distributed throughout the entire network, thus making it
a challenging task. Further, the non-centrally controlled diverse hardware and software 
network infrastructure, allows many additional vulnerabilities open to exploitation by adversaries. A recent example 
features the exploitation of misconfigured NTP (network time protocol) servers, that led  to 
one of the largest DDoS attacks ever recorded~\cite{Czyz:2014:TPG:2663716.2663717}. 
In such {\em reflection and amplification} attacks~\cite{AmplificationHell},
multiple small requests are sent to several mis-configured NTP servers (or other UDP-based services),
which inflict transmissions of large data amounts to targeted hosts.
Thus, the intended victims get overwhelmed with traffic and temporarily disabled.
Volumetric DDoS attacks are just one possible scenario;
low-volume DDoS activities that rely on traffic sparseness
to avoid detection are also of concern.
It is important to be able to
defend against the {\em DDoS threat model} and detect the onset of such potentially
unpredictable attacks in order to adequately secure the network,
e.g., by filtering (blocking) traffic or deploying security patches to network gear.

%In many circumstances, it is also important to be able to {\em reconstruct} the mechanism and feature of such DDoS (ARE THERE GOOD
%REFS FOR THIS PROBLEM? MAYBE EXPAND A BIT. I FOUND THE FOLLOWING
%Anatomy of a Botnet, by Arbor Networks)

The key to addressing these diverse topics  is the availability of {\em adequate data} coupled with {\em advanced monitoring and analysis tools}
and the corresponding {\em software infrastructure}.
%\\
%\noindent {\it ``If we have data, let's look at data. If all we have are opinions, let's go with mine."} -- Jim Barksdale, former Netscape CEO.\\
The vast volume of network traffic streams makes collection, storage and processing of all traffic data infeasible. 
%Further, important ethical and privacy considerations make the storage and analysis of  such data (or significant parts) prohibitive. 
Therefore, the focus has been on the information available in packet headers,
such as source and destination addresses, application ports, protocol, payload size, etc. 
While such type of meta-data is more manageable, its rate of occurrence is still very fast. For 
example, storing packet header information (say initial 96 bytes of an Ethernet frame) 
from a 10 GE link at Merit\footnote{Merit Network, Inc. operates 
Michigan's research and education network. It is an Internet service provider that serves a population of nearly 1 million users.
Merit is the largest IP network in Michigan, and
its network includes a wide range of link types that include link speeds from T1 through 100Gbps. 
The network backbone consists of a 100G fiber ring, which passes through the 
major cities of Michigan, as well as Chicago.}
 at a rate of 1.8 Mpps (million packets per second) 
requires $1.7$ GB per 10 seconds (equivalently, around $15$ TB per single day). 
The industry has developed tools such as NetFlow and others (sFlow, etc.), 
which effectively compress the packet meta data by grouping them into flows.
NetFlow-alike traffic sampling functionality is available on many network elements.
This compression mechanism, however, creates an intermediate step, which introduces a delay
in the access to traffic meta-data (in addition to distorting its structure).   

Even if one has access to  raw data on packet headers or NetFlow, its high acquisition rate makes online analysis of this
information often a formidable challenge.
Many conventional statistical methods and algorithms are not sequential in nature 
and require access to large batches of data spanning several minutes to hours. Thus,
possible DDoS attacks or changes in the network traffic patterns will be 
detected with offline analysis several minutes after their onset.  The time-scales of such analyses are not desirable,
if the goal is to prevent large-scale network outages. Note that 
specialized, albeit very expensive, appliances exist (e.g., Arbor Networks' PeakFlow),
but in real-world settings are configured to receive heavily sampled Flow data (i.e., sampling rates of 1:1000 or more). Hence, low-volume or short-term attacks may elude detection.
Further, such tools require {\em a priori} knowledge of baseline traffic patterns.

These challenges motivate us to develop new software and algorithmic infrastructure for 
harvesting and monitoring network traffic data {\em at the time-scale of routing} (i.e., at wire-speed), guided by the following
principles:
(a) examine {\it all} packets at the monitoring host;
(b) develop memory efficient data structures and statistical summaries that can be computed and retained
{\it at the time-scale of routing};
(c) easy to build and deploy  using {\em commodity, inexpensive, off-the-shelf} hardware;
(d) the resulting data products should be available to be communicated and shared {\it in real-time}
to centralized monitoring stations for further forensics, and
(e) the monitoring architecture should allow for interactive filtering {\it in real-time}.
%\medskip
%{\bf Related work:}
%\vspace{-10pt}
%\subsection{Related Work}

\noindent\textbf{Related Work:}
Over the past 15 years, 
many practical tools have been developed for intrusion detection.
For example, 
\emph{Snort} (see, {\tt snort.org}), \emph{Suricata} ({\tt suricata-ids.org})  and \emph{Bro}
({\tt bro.org}) are popular tools that rely on
\emph{signature-based} methods to examine traffic data for
\emph{known} malicious patterns. Nevertheless, 
recent malware often manage to evade pattern matching detection
by becoming \emph{polymorphic} (i.e., existing in various forms via encryption).
The proposed work aims to complement existing tools by adopting instead 
a \emph{behavioral-based} approach. 

%Over the past 15 years, 
%many practical tools have been developed for addressing many of the issues previously outlined.
%For example, 
%\emph{Snort} (see, {\tt snort.org}), \emph{Suricata} ({\tt suricata-ids.org})  and \emph{Bro}
%({\tt bro.org}) are popular tools that rely on
%\emph{signature-based} methods to examine traffic data for
%\emph{known} malicious patterns.
%Recent malware though often manage to evade pattern matching detection
%by becoming \emph{polymorphic} (i.e., existing in various forms via encryption).
There exists a noteworthy amount of literature in the area of statistical, behavior-based
anomaly detection. Standard techniques
that seek `change detection' points in  traffic time series
include exponential smoothing~\cite{Lucas:1990:EWM:84840.84843}
or other more general time-series techniques~\cite{Box:1990:TSA:574978, Brutlag:2000:ABD:1045502.1045530}. More recent methods
employ wavelet-based tools~\cite{Barford:2002:SAN:637201.637210} or subspace reduction methodologies
based on principal component analysis~\cite{Lakhina:2004:DNT:1030194.1015492, Lakhina:2005:MAU:1080091.1080118}. Such methods lack
the capability of identifying the actual `heavy-hitters'
and, most importantly, suffer
from the `dimensionality curse' (i.e., having multi-dimensional features to monitor)
and/or are inadequate for \emph{online} realization on fast,
multi-gigabit streams. 
Hence, there has  been a lot of activity in the theoretical computer 
science community on designing and studying efficient algorithms for data streams % that address certain aspects of traffic monitoring and anomaly detection problem
that aim to alleviate the high dimensionality and high `velocity' constraints.
Many summary data structures (i.e., sketches) have 
been developed to address the challenging problems
of \emph{identification} of heavy-hitters or frequent items in a
stream~\cite{Cormode1061325, Karp:2003:SAF:762471.762473, Cormode:2003:FHH:1315451.1315492, Estan:2002:NDT:964725.633056, 4146856, Zhang:2004:OIH:1028788.1028802,
 Cormode:2005:IDS:1073713.1073718},
 anomaly \emph{detection} in  high-dimensional regimes~\cite{Li:2006:DIN:1177080.1177099, Krishnamurthy:2003:SCD:948205.948236, 7140420, Zhang:2004:OIH:1028788.1028802},
compressed sensing and estimation of frequency moments~\cite{Gilbert06algorithmiclinear, 
Gilbert:2007:OSF:1250790.1250824, Bar-Yossef:2002:CDE:646978.711822, Alon:1996:SCA:237814.237823, 4385788, 1614066, Indyk:2008:ECC:1347082.1347086},
community mining~\cite{Cormode:2005:SEM:1065167.1065201}, etc. (see~\cite{WICS:WICS1347} and references therein).

In reality, the mere access to fast data streams involves formidable technical challenges.
Many of the more sophisticated  sketch-based algorithms (e.g.,~\cite{Li:2006:DIN:1177080.1177099, Gilbert:2007:OSF:1250790.1250824, 
Gilbert06algorithmiclinear, Indyk:2008:ECC:1347082.1347086}) 
tackle the multi-dimensional aspect of the problem, but
implementing them on multi-gigabit streams 
is rather challenging or often impossible without the use of specialized hardware (e.g., FPGA). 
In addition, few frameworks (e.g.,~\cite{Zhang:2004:OIH:1028788.1028802, Li:2006:DIN:1177080.1177099})
take a holistic approach to develop methods that address the problems 
of change detection and identification together.
Nevertheless, we acknowledge the presence of considerable previous work on the topics of network monitoring, troubleshooting and
intrusion discovery. At the same time, our open-source platform
offers a novel extendible framework that \emph{couples together the important
problems of detection, identification and visualization of aberrant behavior in multi-gigabit streams}.
We propose \emph{new} algorithms that are a direct consequence of the data products generated by our framework.
Furthermore, there are relatively 
few tools that could allow network engineers to interactively examine 10Gbps+ traffic
streams on the time scale of routing.  This motivates the approach we have adopted in this paper, which
focuses on leveraging several recent advances in high-speed packet capture to provide 
tools that are easy and inexpensive to deploy, examine every packet at the interface, 
and provide simple statistics of the `signal' that allow network engineers to {(a)} visualize
structural aspects of traffic, {(b)} detect changes in 
intensity or structure of traffic sub-flows, {(c)} potentially filter and 
zoom-in on anomalous IP address ranges  identified automatically and {(d)} identify  ranges of exact IP addresses associated with anomalous events.  
Table~\ref{tab:related} highlights our contributions and identifies differences with previous literature.

\begin{table}[t!]
\centering
\caption{Characterization of previous work (representative sample).}
\vspace{-15pt}
\footnotesize
\label{tab:related}
\begin{tabular}{r|ccccccc}
       &        &        & Real-time & High-Dim. & Real-time    & Interactive   & Attack         \\
       & Identification & Detection & Visual.   & Data      & Streams      & Zoom-in       & Classification \\\hline
\textbf{AMON}   & \textbf{yes}    & \textbf{yes}    & \textbf{yes}       & \textbf{yes}       & \textbf{yes}          & \textbf{yes} (ongoing work) & \textbf{no}            \\
CGT~\cite{Cormode1061325}    & yes    & no     & no        & yes       & yes          & no            & no             \\
Defeat~\cite{Li:2006:DIN:1177080.1177099} & yes    & yes    & no        & yes       & no           & no            & yes            \\
\cite{Zhang:2004:OIH:1028788.1028802}    & yes    & yes    & no        & yes       & yes         & no            & no             \\
\cite{Lakhina:2005:MAU:1080091.1080118}   & no     & yes    & no        & no        & no           & no            & yes            \\
\cite{7140420}  & no     & yes    & no        & no        & no\tablefootnote{As noted in~\cite{7140420},  it is limited to relatively low rates in an attempt to not overload the device and
affect forwarding actions.} & no            & no            
\end{tabular} 
\vspace{-30pt}
\end{table}

%\vspace{-10pt} 
%\subsection{Major contributions} 
\noindent\textbf{Major Contributions:}
Our first contribution is the design and implementation of a software monitoring framework, referred to as {\em AMON (All-packet MONitor)}, working reliably over 10Gbps+ links. 
This framework is based on {PF\_RING}~\cite{Deri_improvingpassive} in zero-copy mode
which efficiently delivers packets to the monitoring application by bypassing
the OS kernel. We implement hash-based traffic summaries, which simply randomly assign source and destination pairs to bins providing an
aggregate but essentially instantaneous picture of the traffic, which can be used to \emph{diagnose and visualize} 
changes in structure and intensity. 
%We also run parallel instances of Boyer-Moore majority vote
%algorithms~\cite{boyer_moore1991, kallitsis14hashing} for each bin, 
%which with high probability identify {\em exactly} the IP source and destination pairs of `heavy' flows.  
 
Our second contribution is a suite of statistical tools for  automatic \emph{detection} of significant
changes in the structure and intensity of  traffic. 
%While much research on space efficient algorithms is already available, its focus is mostly 
%on providing performance guarantees under certain sparsity or other
%regularity constraints on the data.  Relatively less attention has been payed 
%on the characterization and use of the statistical distribution of traffic flows in the 
%context of anomaly detection in fast data streams. 
The accompanying instances of Boyer-Moore majority vote algorithm~\cite{boyer_moore1991} 
can be leveraged to \emph{identify} precise IP addresses associated with attacks;
this is an important contribution as well.
We illustrate the new data acquisition framework
with several views of the resulting data structures
(hash-binned arrays), referred as \emph{`databricks'}. We show (with real data from Merit!) that even basic visualization tools and algorithms
{\em applied to the right type of data} can help instantaneously identify distributed attacks, which do not contribute to large traffic
volume (see `SSDP' and `Tor' case studies, Section~\ref{subsec:tor}).

This paper is organized as follows: Section~\ref{sec:architecture}
introduces our monitoring architecture, including the {\em data products} (Figure~\ref{fig:data_products})
and our software prototype; Section~\ref{subsec:bm} introduces  {AMON}'s
identification component; 
Section~\ref{sec:stats} discusses
our statistical methodology for automatic detection.
Section~\ref{sec:peva} evaluates our software and algorithms on a rich set
of real-world Internet data, including \emph{four} real DDoS case studies,
and compares against successful and robust state-of-the-art methods for detection
and identification~\cite{Li:2006:DIN:1177080.1177099, Cormode1061325}.

%\begin{figure}[t!]
%\centering
%\includegraphics[width=0.5\textwidth]{./figures/amon_high_level_architecture_v2.pdf}
%\vspace{-15pt}
%\caption{\footnotesize{High-level architecture of AMON.}}
%\label{fig:amon}
%\vspace{-20pt}
%\end{figure}
 
 %\begin{figure}[t!]
 %       \centering
%		\includegraphics[width=0.45\textwidth]{./figures/zc_mon_performance.pdf}
%\vspace{-20pt}
%\caption{\footnotesize{Software performance (Chicago site); rates exceeded 20Gbps, but minute drop rates recorded. }}
%\label{fig:pfring_chicago}
%\vspace{-10pt}
%\end{figure}

\begin{figure*}[t!]
        \centering
        \begin{subfigure}[b]{0.5\textwidth}
              \includegraphics[width=1\textwidth]{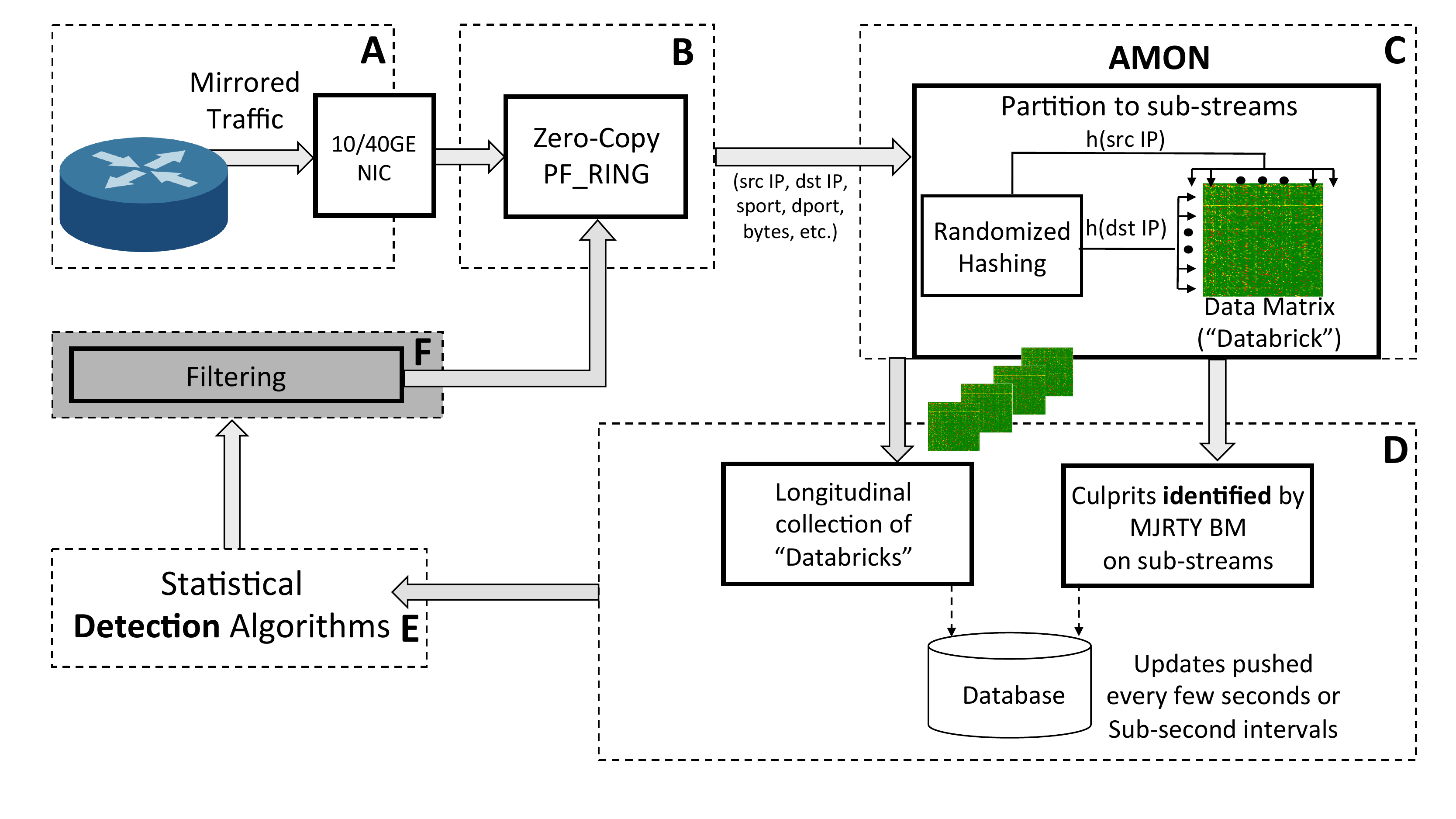}
              %\vspace{-10pt}
              \caption{\footnotesize{High-level architecture of AMON.}}
              \label{fig:amon}
        \end{subfigure}
        \begin{subfigure}[b]{0.45\textwidth}
                \includegraphics[width=0.85\textwidth]{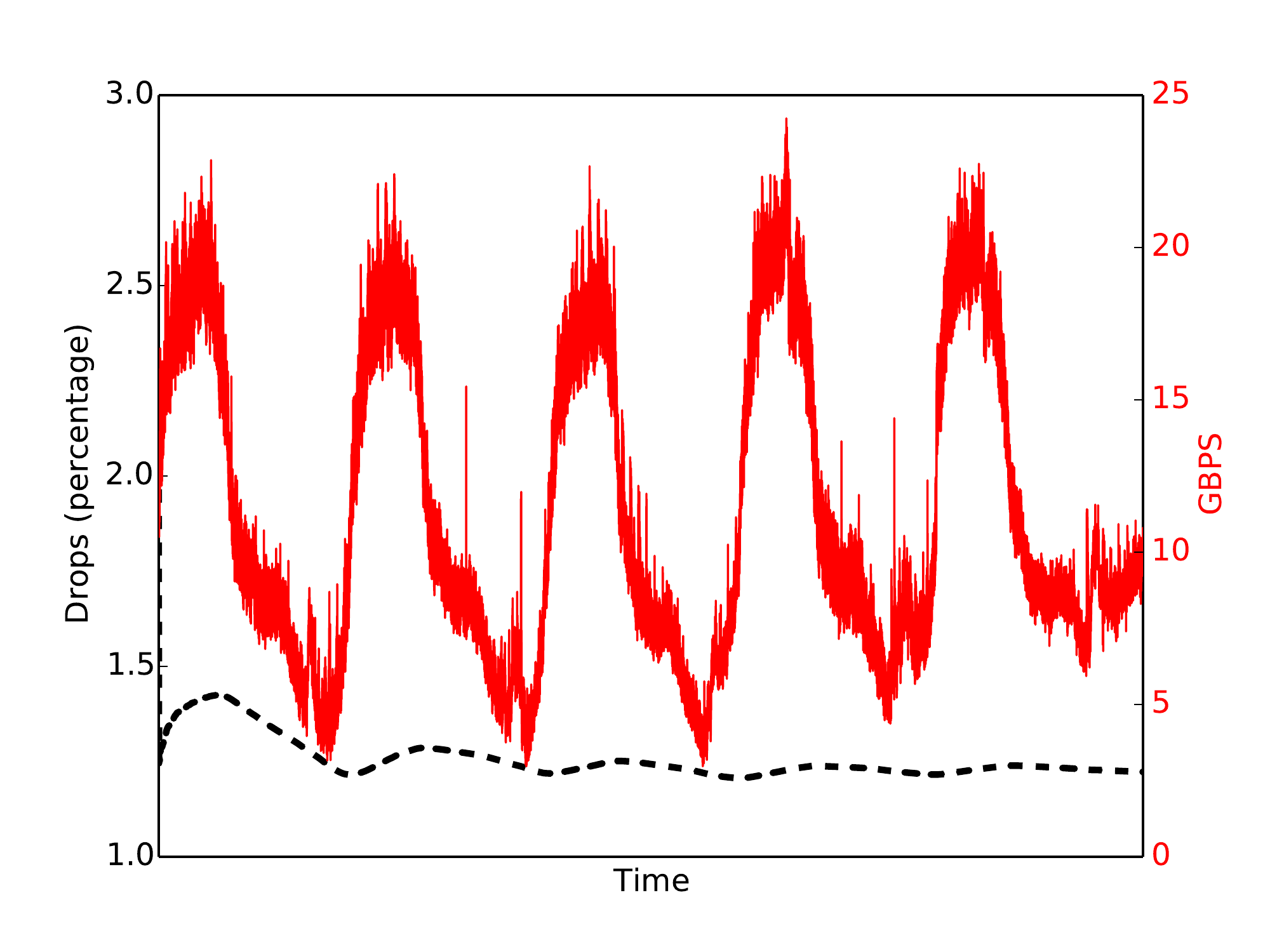}
                 % \vspace{-10pt}
                \caption{\footnotesize{Software performance.}}
                \label{fig:pfring_chicago}
        \end{subfigure}
          \vspace{-15pt}
 \caption{\footnotesize{The proposed framework. \emph{Left:}  AMON's data products comprise the input of identification, statistical detection and
 visualization modules introduced in this paper. \emph{Right:} Performance at rates exceeding 20Gbps; minute drop rates recorded.}}
\label{fig:amon_pfring}
 \vspace{-25pt}
\end{figure*}

\vspace{-7pt} 
\section{Data and software infrastructure}
\vspace{-7pt}
\label{sec:architecture}
An  overview of the proposed architecture is portrayed  in Figure~\ref{fig:amon}.
The monitoring application is installed on a machine that receives raw packets in a streaming fashion.
In our prototype at Merit Network, the monitoring probe receives traffic via a passive traffic mirror (using a SPAN---switched port analyzer---setting) configured on a network switch. Packets are then efficiently delivered at 10Gbps+ rates
to the monitoring module via PF\_RING ZC. 
Subsequently, all packets are processed in a streaming fashion
for constructing, via efficient hashing, a data matrix (i.e., the databrick depicted at Figure~\ref{fig:data_products}) \emph{and}
a separate matrix containing the most active source-destination flows 
identified via our extension of the Boyer-Moore algorithm (Section~\ref{subsec:bm}).
Periodically, these data products are shipped to a database
for storage, further analysis and dashboard-based visualizations.  These data are analyzed 
through various detection algorithms described in Section~\ref{sec:stats}. Flows flagged by the detection module can be extracted
for further analysis by the corresponding filtering mode that is currently under development.
% and would leverage hardware-based techniques performed at the network interface card. 

\vspace{-10pt}
\subsection{Data products via pseudo-random hash functions}
\vspace{-5pt}
Internet traffic monitored at a network interface can be viewed as a stream of items $(\omega_n,v_n),\ n=1,2,\ldots$ 
(see e.g.~\cite{Muthukrishnan:2005:DSA:1166409.1166410}). The $\omega_n\in \Omega$ are the {\em keys} and 
$v_n$ are the 
updates (e.g., payload) of the stream signal. For example, the set of
keys could be all IPv4 addresses ($\Omega = \{0,1\}^{32}$); IPv6 addresses;
pairs of source-destination IP addresses ($\Omega =\{0,1\}^{64}$); 
may include source and destination ports, etc; while payloads could be bytes, packets, distinct ports, etc.
Since it is not feasible to store and manipulate the {\em entire signal} when monitoring 10Gbps+ links, 
we employ \emph{hashing} to compress the \emph{domain} of the incoming stream keys into a smaller set.
Collisions are allowed and, in fact, expected, but the hash function is chosen so that it {\em spreads out} the
set of observed keys approximately uniformly.

\begin{figure*}[t!]
        \centering
        \vspace{-31pt}
        \begin{subfigure}[b]{0.32\textwidth}
               \includegraphics[width=1.1\textwidth, trim=0 160 0 100,clip]{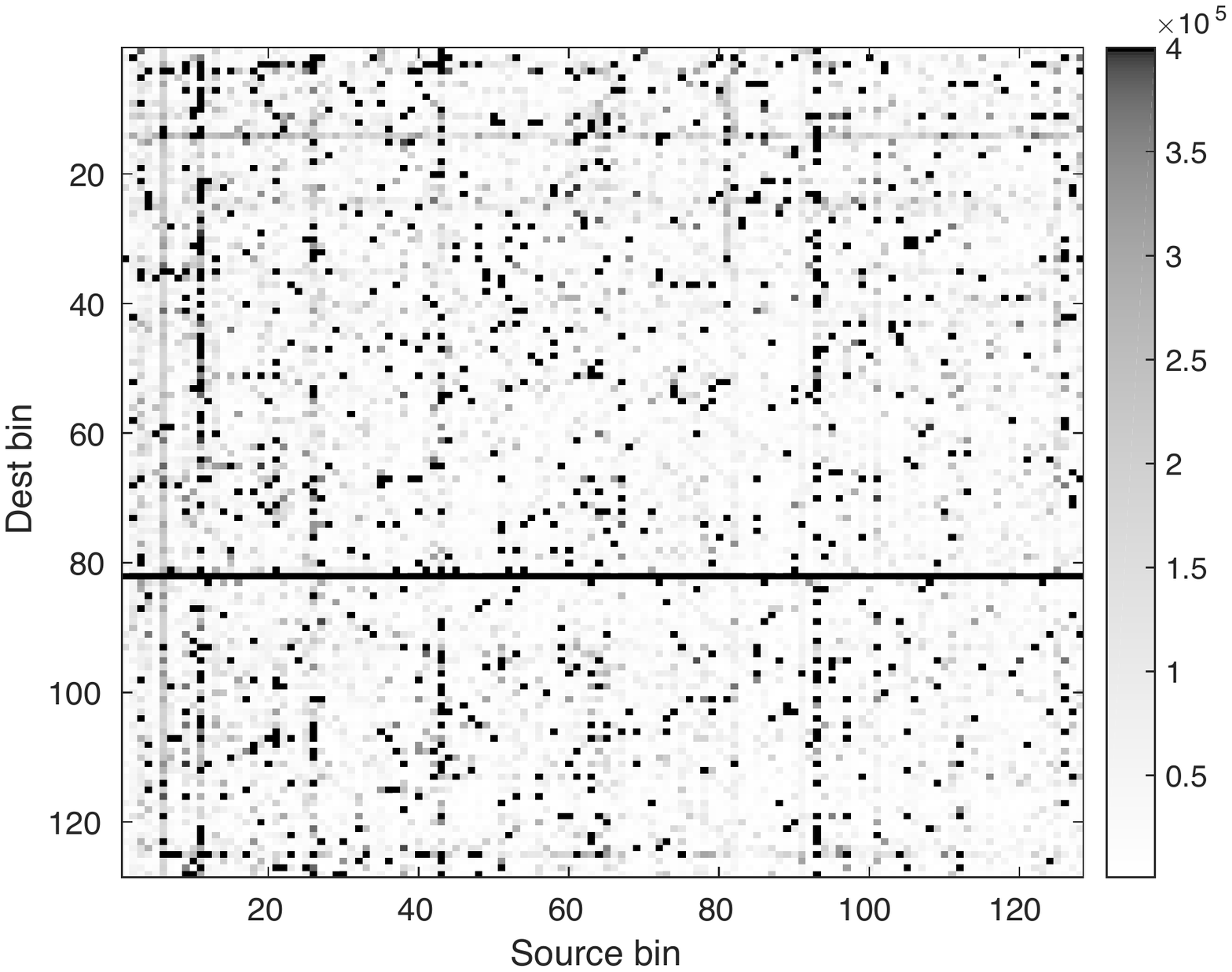}
        \end{subfigure}
        \begin{subfigure}[b]{0.32\textwidth}
                 \includegraphics[width=1.5\textwidth, trim=0 300 0 100,clip]{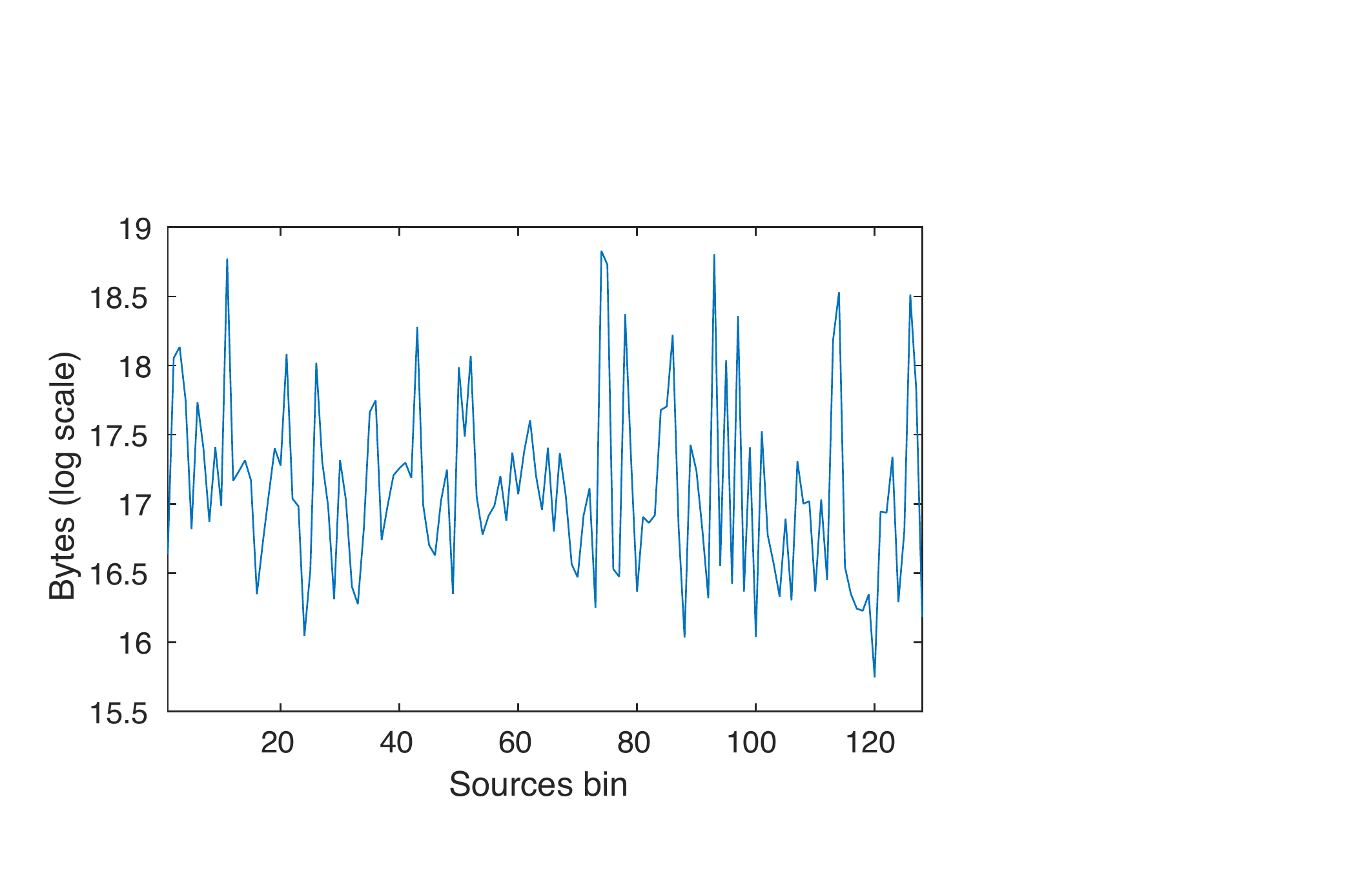}
        \end{subfigure}
        \begin{subfigure}[b]{0.32\textwidth}
                 \includegraphics[width=1.5\textwidth, trim=0 300 0 100,clip]{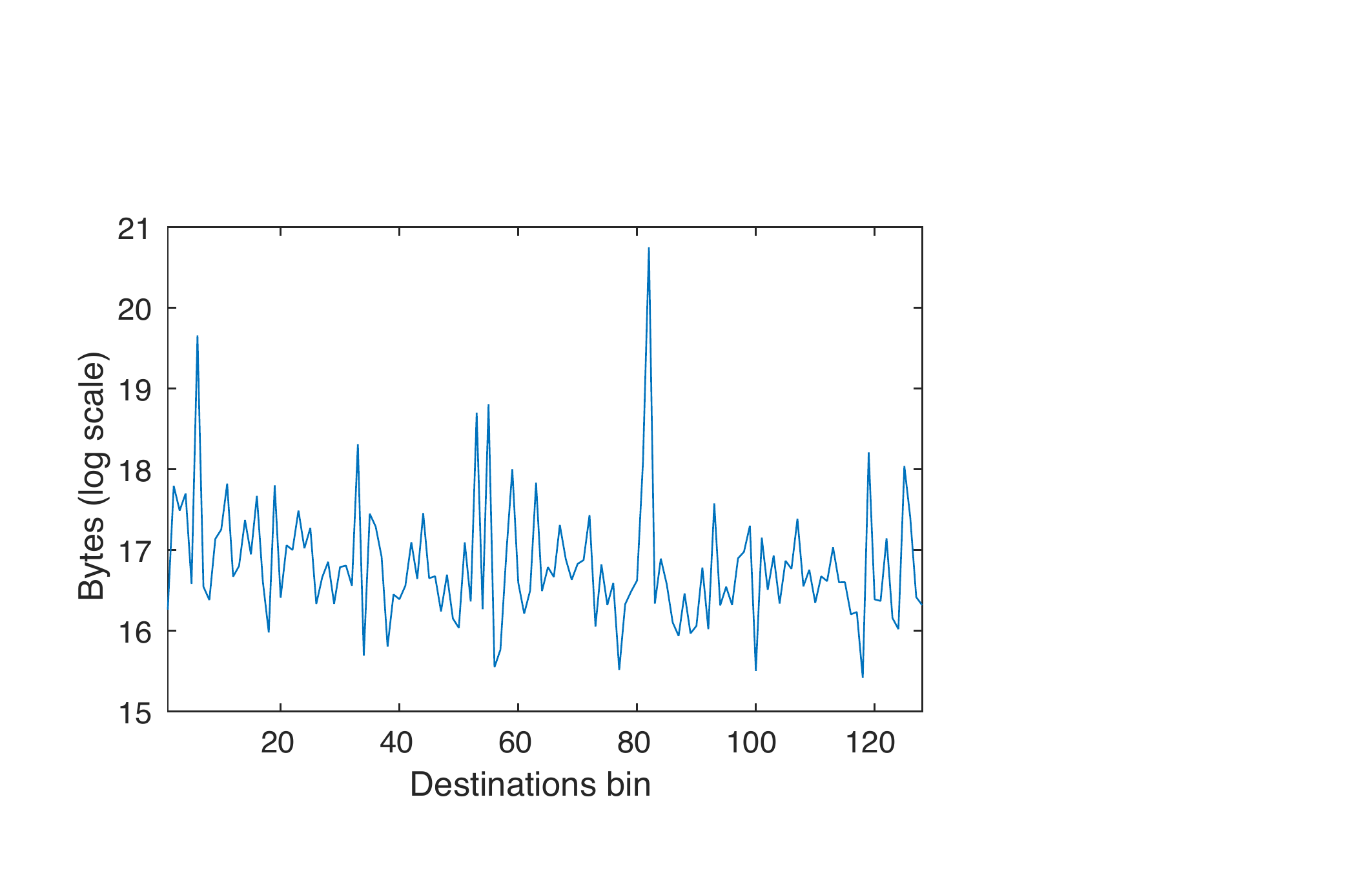}
        \end{subfigure}
        \vspace{-18pt}
 \caption{\footnotesize{Our data products. These  data arrays, generated
         online by our PF\_RING-based software, are used as the basic input structures for our detection algorithms.
         {\em Left:} The `databrick' matrix during the `Library' attack (see Section~\ref{sec:peva});  the apparent 
         horizontal stripe (at dest `bin' 82)  signifies traffic from multiple sources to a single destination (victim). 
         {\em Middle:} View of sources array, constructed by a  matrix column-sum. 
         {\em Right:} Destinations  array; observe that  `bin' 82 stands out (notice the log scale).}}
\label{fig:data_products}
\vspace{-19pt}
\end{figure*}

Consider for example 
the set of  IPv4 addresses $\{0,1\}^{32}$ and
let $h: \{0,1\}^{32} \rightarrow \{1,\ldots,m\}$ be a hash function 
(see~\cite{Carter1979143, kallitsis14hashing, Cormode:2005:IDS:1073713.1073718})
that uniformly spreads the addresses over the interval $\{1,\ldots,m\}$.
Upon observing key $(s,d)$ of the source and destination of a packet, we compute the hashes 
$i:=h(d)$ and $j:=h(s)$ and update the 
\emph{data matrix} $X = \{ X (i,j)\}_{m\times m}$
as 
$
 X(i, j):= X(i, j) + v.
$
This matrix constitutes the first data output of our architecture
and is depicted in Fig.~\ref{fig:data_products}. It is emitted at periodic intervals (e.g., 1 or 10 
seconds) to a centralized database for online as well as further downstream analysis, and reinitialized.
The row- and column-sums of this matrix yield the destination- and source-indexed 
hash-binned arrays, also depicted on the figure. These \emph{data products}
are used as inputs for the detection and visualization algorithms described below.

%\begin{figure*}[t!]
%        \centering
%        \begin{subfigure}[b]{0.32\textwidth}
%               \includegraphics[width=1.0\textwidth]{./figures/Fig_hash_array_databrick-2.eps}
%        \end{subfigure}
%        \begin{subfigure}[b]{0.31\textwidth}
%                 \includegraphics[width=1.0\textwidth]{./figures/Fig_hash_array_sources-2.eps}
%        \end{subfigure}
%        \begin{subfigure}[b]{0.32\textwidth}
%                 \includegraphics[width=1.0\textwidth]{./figures/Fig_hash_array_dests-2.eps}
%        \end{subfigure}
%        \vspace{-10pt}
% \caption{\footnotesize{Our data products. These sketch data blocks, generated
%         online by our PF\_RING-based software, are used as the basic input structures for our detection algorithms.
%         {\em Left:} The databrick matrix; notice the horizontal stripes that signify traffic from multiple destinations to multiple sources. Note also the bold column at source bin 100 that depicts heavy source(s) %activity.  {\em Middle:} View of source array, constructed by aggregating over the columns of the matrix. Note the heavy
%         source(s) at bin 100.  {\em Right:} Destinations  array; observe how heavy destinations appears as heavy bins. In all cases volume is in bytes in log scale.}}
%\label{fig:data_products}
%\vspace{-20pt}
%\end{figure*}

\vspace{-10pt}
\subsection{Software implementation: PF\_RING-based Monitoring}
\vspace{-5pt}
The {AMON} monitoring application is powered by PF\_RING, a high performance packet capture network socket.
Modern hardware advances in CPU speeds and architecture, memory bandwidth and I/O buses
have shifted the bottlenecks in multi-gigabit packet reception into the software 
stack~\cite{Fusco:2010:HSN:1879141.1879169, Gallenmuller:2015:CFH:2772722.2772729}.
PF\_RING avoids unnecessary memory copies between the operating system layers,
and hence the length of the packet journey between the network interface (NIC)
and the monitoring application is shortened.
Consequently, the number of CPU cycles
spent for transferring packets from their NIC entry point to the application
is significantly reduced.
%PF\_RING also efficiently exploits the parallelism offered by multi-core
%computer processors, and the multiple receive (RX) queues offered by modern network adapters. 
This leads to optimal memory bandwidth utilization~\cite{Fusco:2010:HSN:1879141.1879169, Gallenmuller:2015:CFH:2772722.2772729},
and therefore to extremely efficient packet processing speeds (see Figure~\ref{fig:pfring_chicago}). 

Our system takes advantage of the zero-copy framework that PF\_RING offers.
In this mode, the monitoring application reads packets directly from the network interface, i.e., 
both the OS kernel and the PF\_RING module are bypassed.
As a result, efficient monitoring is now achievable with 
commodity, off-the-shelf hardware.
For example, all experiments in this study were conducted using %SuperMicro servers and Intel cards (X540 / X710);
NIC cards costing below 800 USD. 
Although alternative fast packet processing schemes exist~\cite{Gallenmuller:2015:CFH:2772722.2772729}, PF\_RING
was selected due to its robustness, proved efficiency and broad versatility.

\vspace{-15pt}
\section{Identification: the hash-thinned Boyer--Moore algorithm}
\label{subsec:bm}

The proposed architecture periodically emits
a list of heavy activity stream elements that can be used for traffic engineering purposes, accounting and
security forensics. When an alert is raised, 
operators can readily examine these `heavy-hitters'.
 Our identification algorithm is based on the \emph{Boyer-Moore} (BM) majority vote algorithm~\cite{boyer_moore1991},
 and the idea of stream thinning for creating sub-streams described below. The so-named
 MJRTY Boyer-Moore algorithm~\cite{boyer_moore1991} can identify \emph{exactly}
  the majority element---the element whose volume is at least 50\% of the total---in a stream, if one exists.
 It solves the problem in time {\em linear} in the length of the input sequence and {\em constant} memory.
 We first define the identification problem at hand.
%However, in most real settings, a single IP or flow is not responsible for such a high fraction of the total volume of the stream. 
 %Hence, identification accuracy of the plain Boyer-Moore algorithm could be lacking. 
% To overcome this, we employ the Boyer-Moore
% idea on sub-streams of $\cal{S}$.
%In practice, as Figure~\ref{fig:bm_eval} illustrates, thinning using $m=256$  makes the algorithm to perform remarkably well, but one could use a higher $m$ to further thin stream $\cal{S}$, if needed.  
 
 \begin{defn} (\textsc{Identification of Heavy-hitters}) Given an input stream $(\omega_n,v_n)$,
 identify the top-$K$ most frequent items. The frequency of key $\omega$ is the sum of its updates $v$.
 \end{defn}
 
Next, we describe the original  Boyer-Moore algorithm with an analogy to the \emph{one-dimensional}
 random walk on the line of non-negative integers.  A variable $count$ is initialized to $0$ (i.e., the origin)
 and a  candidate variable \emph{cand} is reserved for use. Once a new key arrives,
 we check to see if \emph{count} is 0. If it is, that IP is set to be the new candidate \emph{cand}
and  we move \emph{count} one step-up, i.e. $count=1$. Otherwise, if the IP is the same as \emph{cand}, then 
\emph{cand} remains unchanged and \emph{count} is incremented, and, if not, \emph{count} moves one step-down 
(decremented). We then proceed to the next IP and repeat the procedure. Provably, when all IPs are read,
\emph{cand} will hold  the one with majority, if majority exists.  

Our extension of the MJRTY Boyer-Moore method 
applied to each sub-stream is outlined in Figure~\ref{alg:bm}. 
It returns up to $m$ `heavy-hitter' items present in a stream of keys, taking values in $\{\omega_1, \ldots, \omega_N\}$,
by `thinning'  the original stream $\mathcal{S}$ into $m$ sub-streams. In the \emph{update} operation, upon observing a new stream item $(\omega, v)$,
we compute the sub-stream index $s:=h_1(\omega)$ using hash function $h_1$. In essence, we run $m$
independent realizations of the Boyer-Moore algorithm described above, one for each sub-stream.
Arrays \emph{count} and $cand$ hold the algorithm state for all sub-streams, i.e., $cand[s]$ holds the majority candidate for $s$.
Array $count$ is updated accordingly with the value of $v$ as lines 9, 11 and 16 depict.
The auxiliary ${flag}$ for each sub-stream $s$ 
can help track whether a majority is indeed underlying into that sub-stream; at the start of the monitoring
period the  ${flag}$  corresponding to $s$ is set, and as long as $count[s]$ remains non-negative (i.e., $cand[s]$ needs no updates),
the ${flag}$  never resets. A ${flag}$  that remains `on' guarantees the presence of a majority 
item\footnote{As an example, during the real-time 5-day experiment with the Chicago traffic (Figure~\ref{fig:pfring_chicago}),
an average fraction of at least $85.41\%$ of all sub-streams (with $0.11\%$ standard deviation) contained a majority element (for bytes).}.
An estimate of the 
frequency of each hitter can be obtained via the $m\times m'$ data structure $P_{bm}$.
Through the use of an independent hash function $h_2$, this sketch structure
keeps a hash array of size $m'$ for each sub-stream $s$, and gets updated with the arrival
of each stream element.

When a \emph{query} operation is performed (see Figure~\ref{alg:bm}), we retrieve the $m$ candidates.
An estimation of the volume of a  candidate `hot' item $s$ is recovered by looking at the maximum value
of  sub-array $P_{bm}[s]$. The $m$ candidates are
ranked according to these estimates, and an approximation of the set of top-$K$ hitters is identified.

\begin{figure}
    \begin{subfigure}{.5\textwidth}
     \renewcommand{\algorithmicrequire}{\textbf{Input:}}
    \centering
    \begin{algorithmic}[1]
    \scriptsize 
    \REQUIRE Number heavy-hitters: $K$, $K\le m$
    \REQUIRE hash function $h_1: [N] \to [m]$
    \REQUIRE hash function $h_2: [N] \to [m']$, $m' = 256$
    \STATE $count[i] = 0$, \ $i\in[m]$
    \STATE $cand[i] = -1$, \ $i\in[m]$
     \STATE $flag[i] = 1$, \ $i\in[m]$
     \STATE $P_{bm}[i,j]=0$, \ $i\in[m]$, $j\in [m']$. 
     \end{algorithmic}
     \vspace{-1pt}
     \caption{Initialization}
     \vspace{-125pt}
    \end{subfigure}% need this comment symbol to avoid overfull hbox
    \begin{subfigure}{.5\textwidth}
        %\caption{How to write algorithms}
    \end{subfigure}\\
        \begin{subfigure}{.5\textwidth}
      \centering
       \begin{algorithmic}[1]
       \scriptsize 
         \STATE $P_{est}[s] = \max_{j\in[m']} P_{bm}[s,j]$, \ $s\in[m]$
       \STATE Initialize ${\cal O} = \emptyset$
   \FOR{r=1 \TO K} 
        \STATE Find $o = {\rm Argmax}_{j\in[m] \setminus {\cal O}} P_{est}[j]$ 
        \STATE ${\cal O}  = {\cal O} \cup  o $  {\#Exclude for next iteration }
         \STATE Output $\omega_r^* = $cand$[o]$ 
          \STATE Output $P_{est}[o]$ 
  \ENDFOR
   \end{algorithmic}
        \vspace{-5pt}
        \caption{Query operation}
    \vspace{-161pt}    
    \end{subfigure}\hspace{-30pt}
    \begin{subfigure}{.5\textwidth}
      \centering
      \vspace{-15pt}
     \begin{algorithmic}[1]
       \scriptsize 
\STATE [Thin]  Compute buckets $s=h_1(\omega)$ and $j = h_2 (\omega)$.
\IF{$cand[s] == -1$}
   \STATE $cand$[s] = $\omega$, $count$[s] = $v$, $P_{bm}$[s,j] = $v$
\ELSE
    \IF{$cand[s]==\omega$}
         \STATE $P_{bm}$[s,j] =  $P_{bm}$[s,j]  + $v$, $count$[s] =  $count$[s] + $v$
     \ELSE
          \IF{$count[s] > 0$}
               \STATE $count[s]$ = $count[s]$ - $v$
               \IF{$count[s] <0$}
                    \STATE $cand$[s] = $\omega$, $count$[s] = -$count$[s]   {\# reset cand}
                     \STATE  $flag$[s] = 0 {\# reset flag}
               \ENDIF
               \STATE $P_{bm}$[s,j] =  $P_{bm}$[s,j]  + $v$
          \ELSE 
             \STATE  $cand$[s] = $\omega$,  $count$[s] = $v$ {\# reset cand}
              \STATE  $flag$[s] = 0 {\# reset flag}
             \STATE  $P_{bm}$[s,j] = $P_{bm}$[s,j]  + $v$ 
          \ENDIF   
    \ENDIF
\ENDIF
 \end{algorithmic}
 \vspace{-5pt}
   \caption{Update operation for stream item ($\omega, v$)}
    \end{subfigure}%
    \hspace{30pt}
 \vspace{-10pt}
\caption{Identification algorithm: Hash-thinned MJRTY Boyer-Moore.}
\label{alg:bm}
\vspace{-30pt}
\end{figure}

\vspace{-17pt}
\section{Statistical Methods for Anomaly Detection}
\label{sec:stats}
\vspace{-7pt}
 This section introduces three new detection methods. We start with a data exploration and model
 validation discussion that characterizes our data products;
 all methods leverage AMON's data. The first method
 is based on estimating the \emph{number of `heavy hitters'} at each time point; 
 this estimate represents a \emph{monitoring statistic} that one can track, and 
 time points with heavy hitter activity can be flagged as anomalies. Next, a method derived by modeling 
 the distribution of the \emph{relative volume} of the heaviest bins in the  source and destination, 1-dimensional, hash-binned arrays is introduced. This
 section concludes with a technique for discovering \emph{structural changes} in traffic, a method
 chiefly suitable for seemingly innocuous, low volume aberrant behavior. Henceforth, the problem
 of anomaly detection is formulated as follows.
 
 \begin{defn}{(\textsc{Detection})} Given an input traffic stream, find the time points
 when the baseline probability distribution of an appropriate monitoring statistic seems inadequate.
 \end{defn}
 
 \vspace{-20pt}
 \subsection{Statistics, model validation and data exploration}
 \label{sec:vis}

\begin{figure*}[t!]
        \centering
        \begin{subfigure}[b]{0.48\textwidth}
               \includegraphics[width=1\textwidth]{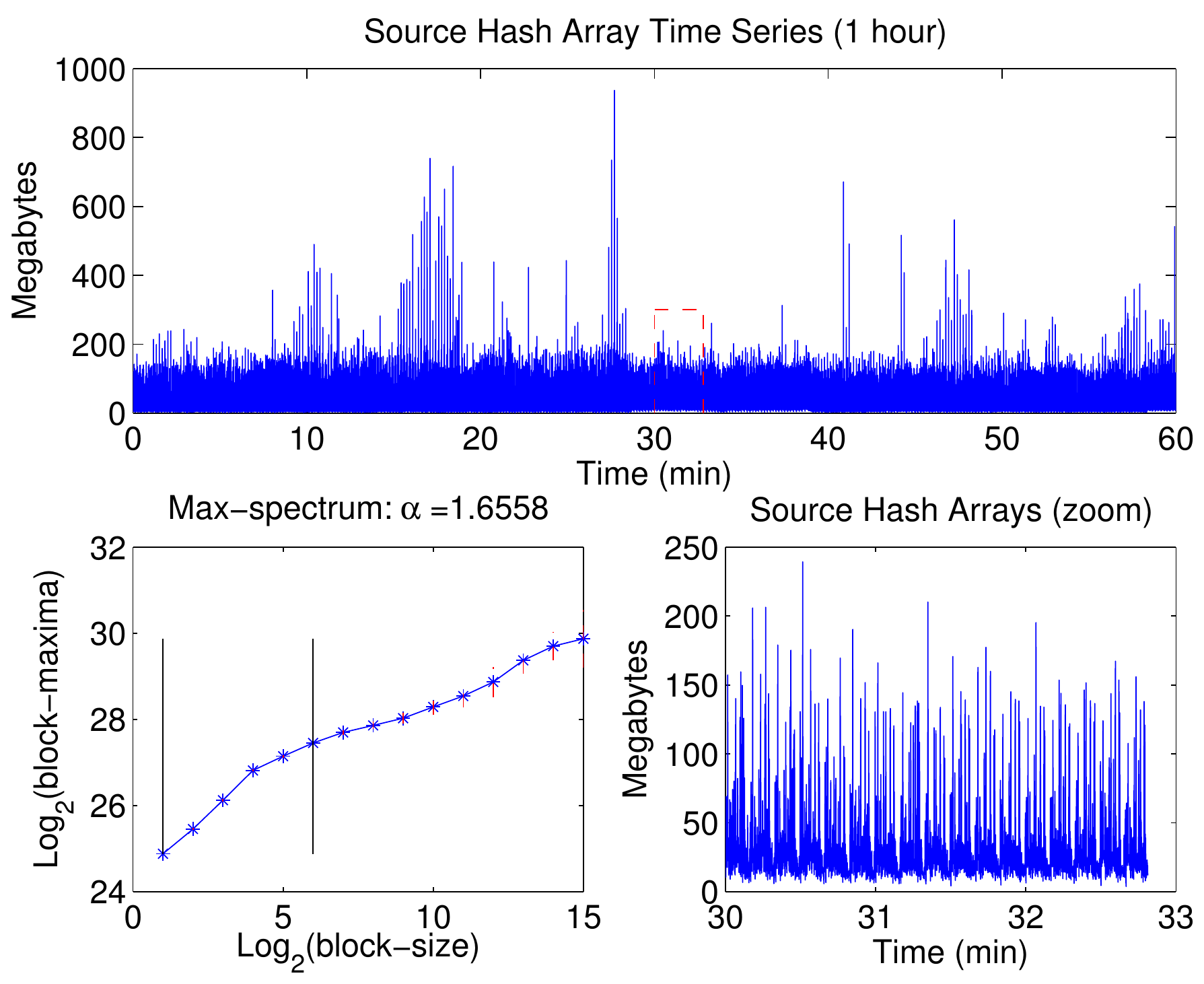}
        \end{subfigure}
        \begin{subfigure}[b]{0.50\textwidth}
               \includegraphics[width=1\textwidth]{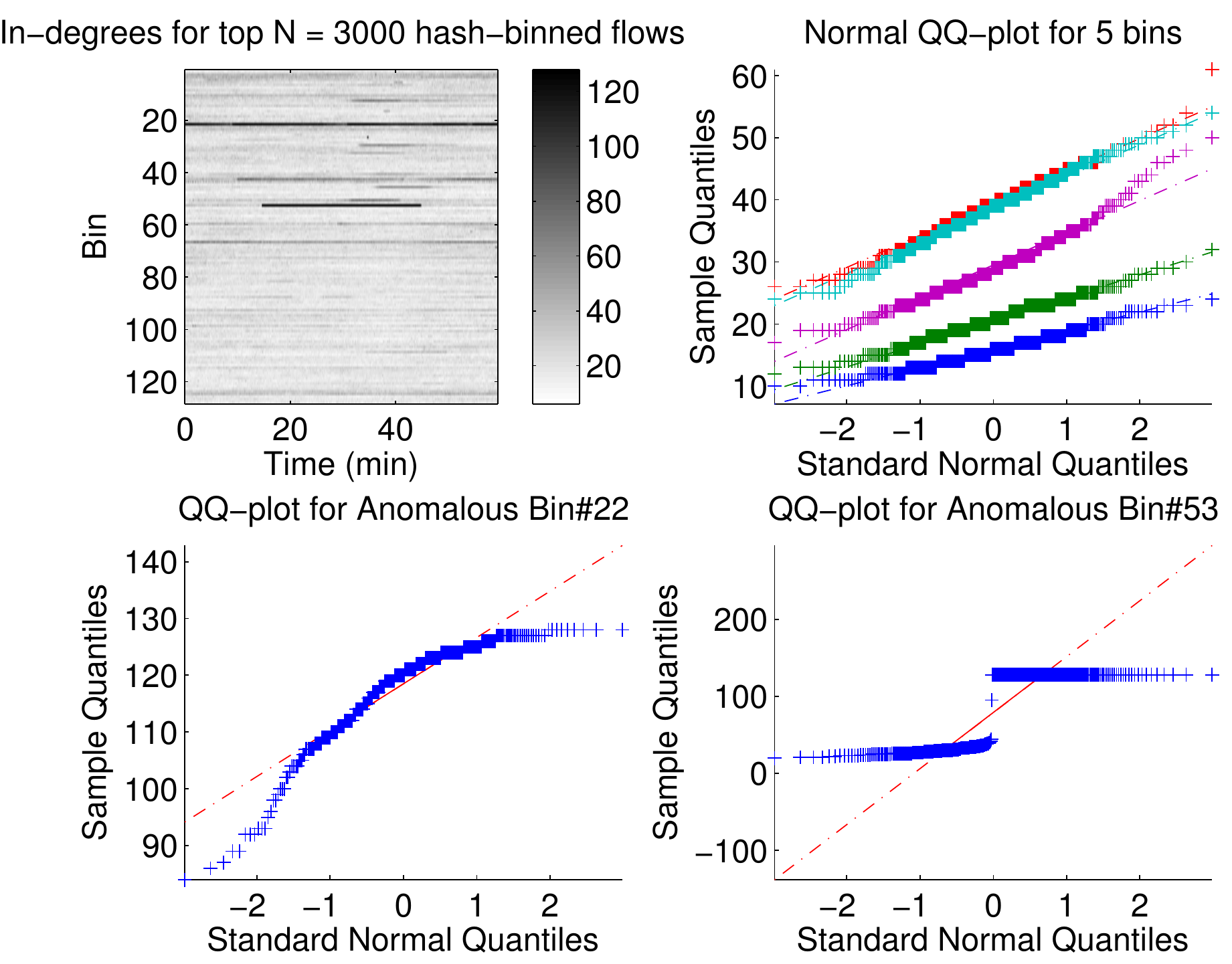}
        \end{subfigure}
 \caption{\footnotesize{{\em Left panel:} Time-series of Source hash-binned arrays (Top) and its zoomed-in version (Bottom-right), computed over 10-second     windows.  The {\em max-spectrum} of the entire time series is plotted on the bottom-left. 
Merit Network: 17:30-18:30 EST, July 22, 2015.
   {\em Right panel:} Merit Network 16:00-17:00 EST, Aug 1, 2015 -- the `Tor' event in Section \ref{subsec:tor}.
(Top-left) Ingress connectivity for the top $N=3000$ hash-binned flows per 10-second windows over 1-hour. (Top-right) QQ-plots demonstrating accuracy of the Normal approximation of typical in-degree distributions. (Bottom plots) QQ-plots for anomalous bins.}}
\label{fig:src_series_conn_dist}
\vspace{-15pt}
\end{figure*}

 The successful detection of statistically significant anomalies in the derived hash-binned traffic arrays depends on the adequacy
of the model employed. We undertook an extensive empirical analysis of long hash-binned arrays and found 
 that heavy-tails are ubiquitous. Figure \ref{fig:src_series_conn_dist} (left panel, top figure) shows a time series of the linearized
 hash-binned array of outgoing (Source) traffic at Merit Network for the period 17:30-18:30 EST 
 on July 22, 2015. %The traffic is from a monitoring link at Merit's Detroit peering point.
 Observe the consistent presence of extreme peaks in the data, some of which
 may in fact be due to an attack event (see the `Library' case study, Section~\ref{sec:peva}). By zooming-in on a short (seemingly calm) $3$-minute period---bottom right---we observe that the extreme peaks, although of lower magnitude, persist.  
% The seeming periodicity in the data is due to the fact that we have concatenated (linearized) hash-arrays $X_t(i)$ over time---this period of about %3 minutes  involves 17 hash-arrays.
 
%\begin{figure}[t!]
%        \centering
%        \includegraphics[width=0.7\textwidth]{./figures/Fig_SRC_time_series_and_max_spectrum.eps}
%\caption{Time-series of Source hash-binned arrays (Top) and its zoomed-in version (Bottom-right), computed over 10-second windows. 
%The {\em max-spectrum} of the entire time series is plotted on the bottom-left. Merit Network: July 22, 2015, 17:30-18:30 EST.}
%\label{fig:src-time-series}
%\end{figure}

%\begin{figure}[ht!]
%\includegraphics[width=0.6\textwidth]{./figures/Connectivity_Distributions.eps}
%\caption{Merit Network 16:00-17:00, Aug 1, 2015 -- the `Tor' event in Section \ref{subsec:tor}.
%{\em Top-left:} Ingress connectivity for the top $N=3000$ hash-binned flows per 10-second windows over 1-hour. {\em Top-right:} QQ-plots %demonstrating
%accuracy of the Normal approximation of typical in-degree distributions. {\em Bottom plots}: QQ-plots for anomalous bins. }
%\label{fig:connectivity-distributions}
%\end{figure}

Heavy-tailed power law distributions are suitable statistical models for data exhibiting such characteristics. Power laws are ubiquitous 
 in computer network traffic measurements. It is well known and documented that file-sizes, web-pages, Ethernet traffic, etc.\ exhibit 
power-law tails~\cite{282603,650143}. Specifically, let $X = X_t(i)$ denote, for example, the amount of traffic registered in a given hash-array bin 
$i$.  Then, a parsimonious model for its tail is as follows:
\begin{equation}\label{e:Pareto-tail}
\P( X >x ) \sim c/x^{\alpha}, \ \ \mbox{ as } x\to\infty,
\end{equation}
where `$\sim$' means that the ratio of the left- to the right-hand side converges to $1$ 
and where $\alpha>0$ and $c>0$ are constants. The smaller the exponent $\alpha$,
the heavier the tail of the distribution, and the greater the frequency of extreme values. In particular, if $\alpha<2$, then the variance 
of $X$ does not exist and if $\alpha<1$, then the mean of this model is infinite.

Figure~\ref{fig:src_series_conn_dist} (left panel, bottom) shows the {\em max-spectrum} of the entire 1-hour long time series of hash-binned 
source traffic array. The {\em max-spectrum} is a plot of the mean log-block-maxima versus the log-block-sizes of the data.  A {\em linear} trend 
indicates the presence of power-law tails as in \eqref{e:Pareto-tail}, while the slope provides a consistent estimate of $1/\alpha$.  Thus, steeper
max-spectra correspond to lower values of $\alpha$ and heavier tailed distributions generating more extreme values. A useful 
feature of the max-spectrum plot is its ability to examine  various log-block-sizes (scales), 
thus enabling simultaneous  examination of  the 
power-law behavior in  the data at various time-scales~\cite{stoev:michailidis:taqqu:2011}. 
As it can be seen, the power-law behavior (linearity in the spectrum) extends over a wide range of time-scales from seconds to hours.   
The time-scale relevant to our studies is a few seconds, which yields estimates of $\alpha \approx 1.6$, obtained by fitting a line 
over the range of scales ($\log_2$-block-sizes) 1 through 6.  Over intermediate time-scales (a few minutes) the  exponent 
$\alpha$ raises to about $2.5$. The simple power-law models are no longer sufficient to capture the distribution 
over the largest time-scales (hours), where complex intermittent non-stationarity and diurnal trends dominate.
%Figure 5 of our extended version~\cite{kallitsis15amon} (see also discussion therein) gives further
%evidence that over very short time scales of a few seconds to a minute,
%the power-law model captures the essence of the distribution.  

\begin{figure*}[t!]
        \centering
        \begin{subfigure}[b]{0.49\textwidth}
               \includegraphics[width=1\textwidth]{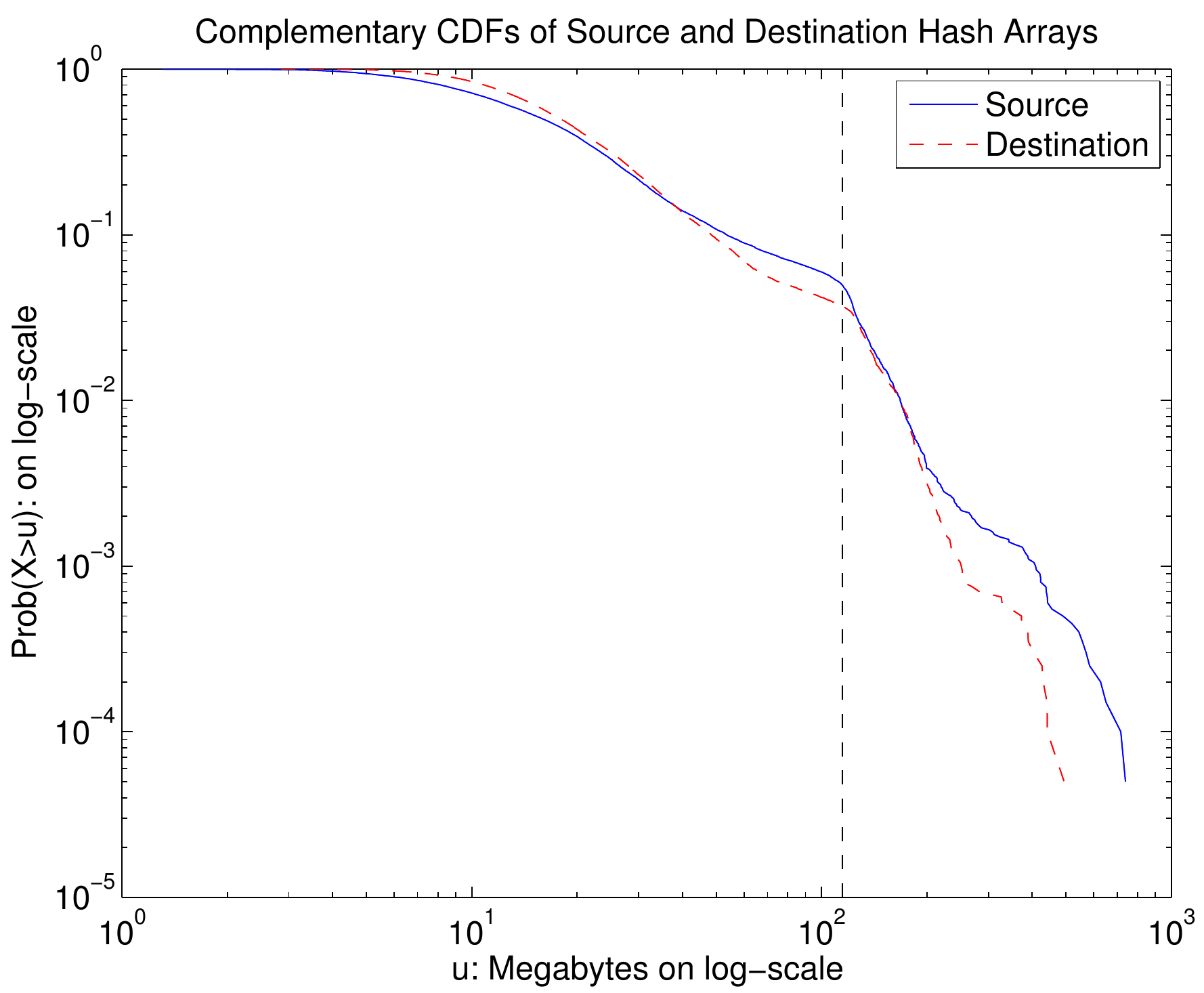}
        \end{subfigure}
        \begin{subfigure}[b]{0.49\textwidth}
                 \includegraphics[width=1\textwidth]{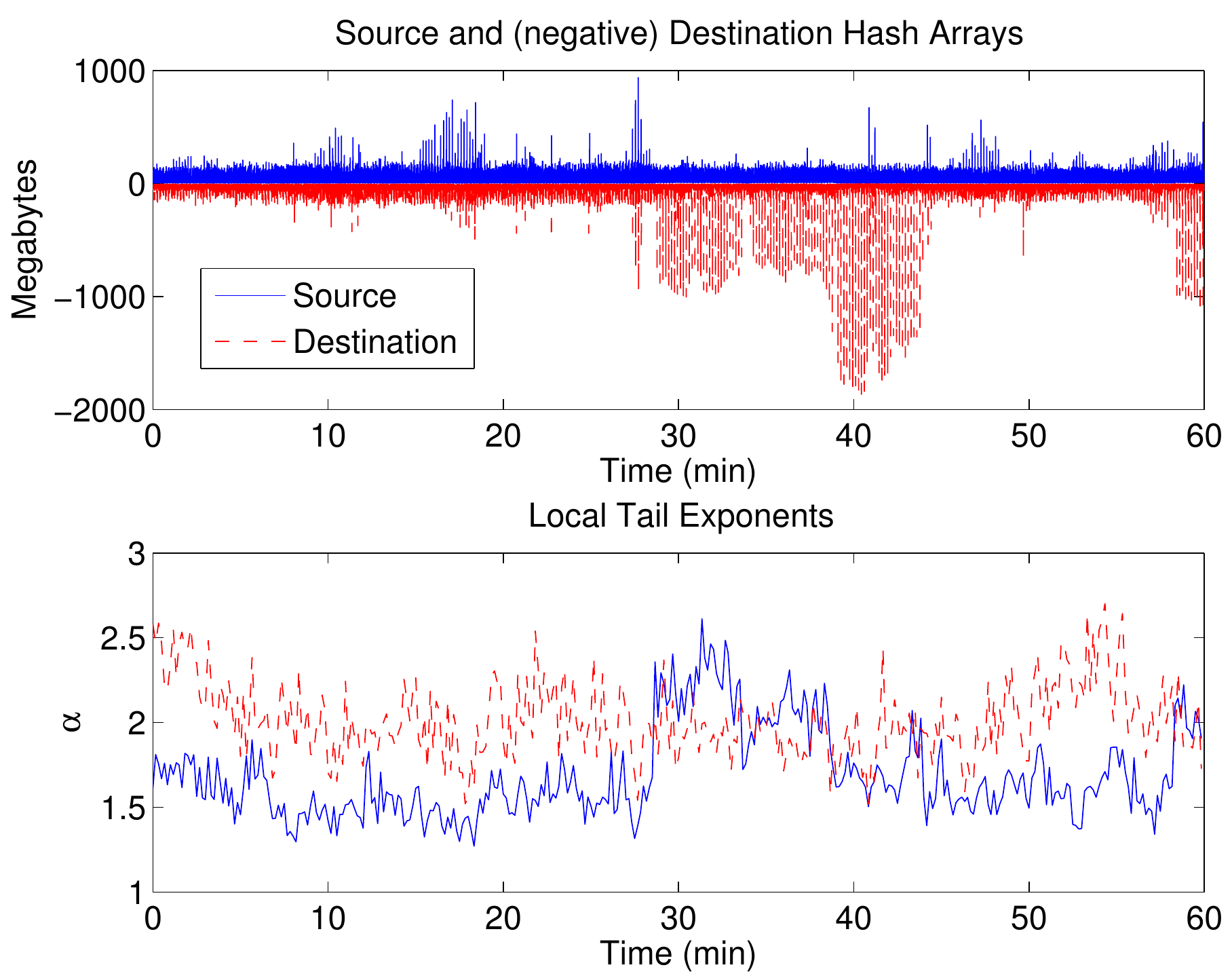}
        \end{subfigure}
 \caption{\footnotesize{{\em Left:} Complementary CDF $u\mapsto \P(X> u)$ on log-scale for Source and Destination traffic hash-arrays.
{\em Right:} Time series and tail exponents of Source and Destination traffic hash-array computed over $10$-second windows. 
 Merit Network: July 22, 2015, 17:30-18:30 EST. }}
 \label{fig:CCDF}
 \end{figure*}

Alternatively, Figure~\ref{fig:CCDF} shows the complimentary cumulative distribution function $x\mapsto \P(X>x)$ on a 
log-scale for both Source and Destination traffic arrays. Linear scaling on this plot corresponds to power-law behavior as in 
\eqref{e:Pareto-tail} and the slope of the linear fit yields an estimate of $-\alpha$. Even though one cannot clearly talk about 
time-scales here, similarly to the max-spectrum plot, one sees two regimes of power-law scaling. The heavy-tail behavior is 
relatively more severe for the range of smaller values corresponding, on average, to shorter time-scales.
Our focus is on very short time scales of a few seconds to a minute.  Our analysis shows that over such time-scales, 
the power-law model captures the essence of the distribution.  Figure~\ref{fig:CCDF} shows tail exponents for both 
Source and Destination hash arrays $X_t(i),\ i=1,\ldots,m$ as a function of time $t$. Observe the persistent heavy-tailed 
nature of the data throughout the entire period of time. Note that the Source (outgoing) traffic is slightly heavier-tailed 
(lower exponents $\alpha$) than the Destination (incoming).  Note also that the estimators of the tail exponent are rather robust to
large-volume fluctuations, e.g.\ in the Destination time-series. This is another important feature of the max-spectrum, which will play a 
role in the successful detection of such anomalous event, described in the following sections.

\vspace{-15pt}
 \subsection{Detection of Heavy Hitters}
 \vspace{-7pt}
 \label{sec:detecthh}
 \label{sec:detection}
In this section, we describe a methodology (named as `Fr\'echet method') for 
aberrant behavior discovery, based on monitoring the number of 
hash-bins involving heavy traffic, henceforth referred as  {\em heavy hitters}.  
%There is extensive literature on the estimation of `heavy hitters' in fast network traffic 
%streams~\cite{Karp:2003:SAF:762471.762473, Cormode:2005:IDS:1073713.1073718, Cormode:2003:FHH:1315451.1315492,
%Estan:2002:NDT:964725.633056, Gilbert06algorithmiclinear, Krishnamurthy:2003:SCD:948205.948236, 4146856, 1354567, Cormode1061325,  Porat:2012:STM:%2095116.2095212, doi:10.1137/100816705}.  
The precise definition of a
heavy hitter is rather subtle; depending on the traffic context, a given flow (e.g.\ video transmission) may be perceived as a heavy hitter in light traffic
conditions, while it may be in fact a `typical' event in normal traffic conditions. Here, we adopt a statistical perspective, where we flag hash-bins as heavy hitters, if 
their signal exceeds a given quantile of a \emph{baseline probability distribution}, i.e., we view heavy hitters as `outliers'.  In order to be adaptive to changing traffic 
conditions, we shall dynamically and robustly estimate the baseline probability model from the data. Our notion of heavy hitters depends on the probability 
associated with the quantile threshold.  This tuning parameter may be set depending on how sensitive we would like to be to `alarms'. 
%We now provide more details.

Let $X_t = \{X_t(i)\}_{i=1}^m,\ t=1,2,\cdots$ be sequence of hash-binned traffic arrays. In the case of the {\em source IP} signal, for example, $X_t(i)$ corresponds
to the column-sum of the databrick matrix and represents the 
 number of bytes originating from all source IPs $\omega$ hashed to bin $i$, i.e.\ $h(\omega)=i$ over the time-window $t$. 
Figure \ref{fig:data_products} (middle) shows  an example of such array.
% based on zero-copy PF\_RING packet tap of a 10GE link at Merit. Observe the rather heavy-tailed nature of the hash-binned traffic,
%discussed extensively in Section \ref{sec:vis}. 
The process of hashing effectively randomizes traffic flows in different bins, and therefore, the entries 
 $X_t(i),\ i=1,\ldots,m$ may be reasonably assumed to be statistically independent and identically distributed (i.i.d.). 

Our goal is to identify and flag the presence of abnormally large (heavy) traffic.  One way that this manifests itself is through 
abnormally large values of $X_t(i)$s, for {\em some} $i$'s. To this end, we consider the sample maximum of the hash-array:
\begin{equation}\label{e:D}
  D_m(X_t):= \max_{i=1,\ldots,m} X_t(i).
\end{equation}
We shall identify a bin $i\in\{1,\ldots,m\}$ as a heavy hitter, if its value is large, relative to an {\em asymptotic approximation} to the
distribution of the sample maximum.  

\begin{prop}\label{p:Frechet-limit} Let $X(i),\ i=1,\ldots,m$ be i.i.d.\ random variables with heavy tails as in \eqref{e:Pareto-tail}.
Then, as $m\to\infty$, we have that
\begin{equation}\label{e:p:Frechet-limit}
\frac{1}{m^{1/\alpha}} D_m(X) \equiv \frac{1}{m^{1/\alpha}} \max_{i=1,\ldots,m} X(i) \stackrel{d}\longrightarrow  c^{1/\alpha} Z_\alpha,
\end{equation}
where $\P(Z_\alpha \le x) = e^{-1/x^\alpha}$ has the standard $\alpha$-Fr\'echet distribution and $c$ is the asymptotic
parameter in \eqref{e:Pareto-tail}.
\end{prop}

This result is a simple consequence of Theorem 3.3.7, p.\ 131 in \cite{embrechts:kluppelberg:mikosch:1997}.
For completeness, its proof is given in the Appendix.
%{\bf MOVE TO THE EXTENDED VERSION.}
%\begin{IEEEproof}
%By the independence of the $X(i)$'s, for all fixed $x >0$, we have
%$$
%\P( m^{-1/\alpha} D_m(X) \le x ) = \P( X\le m^{1/\alpha} x)^m = (1- \P(X> m^{1/\alpha}x))^m. 
%$$
%Now, by \eqref{e:Pareto-tail} with $x$ replaced by $m^{1/\alpha}x$, we observe that 
%$\P(X>m^{1/\alpha}x) \sim c/(mx^\alpha)$, as $m\to\infty$. Thus, using the fact that 
%$(1- c x^{-\alpha}/m)^m \to e^{-c/x^{\alpha}},\ m\to\infty$, we conclude that
%$$
% \P( m^{-1/\alpha} D_m(X) \le x ) \longrightarrow e^{-c/x^{\alpha}},\ \ \mbox{ as }m\to\infty.
%$$
%This implies the desired convergence in \eqref{e:p:Frechet-limit}, since 
%$\P( c^{1/\alpha} Z_\alpha \le x) = e^{-c/x^\alpha},\ x>0.$
%\end{IEEEproof}

Relation \eqref{e:p:Frechet-limit} suggests that for relatively large values of $m$, we can use the limit $\alpha$-Fr\'echet distribution to 
calibrate the detection of heavy hitters.  Specifically, given a sensitivity level $p_0$ (e.g.\ equal to $0.95$ or $0.99$), we flag the bin $i$ as a 
{\em heavy hitter}, if 
\begin{equation}\label{e:Frechet-threshold}
X_t(i) \ge T_{p_0} (m,\alpha,c):= m^{1/\alpha} c^{1/\alpha} \Phi_\alpha^{-1}(p_0) = {\Big(} \frac{c}{\log(1/p_0))} {\Big)}^{1/\alpha},
\end{equation}
where $\Phi_\alpha^{-1}(p) = (\log(1/p))^{-1/\alpha},\ p\in(0,1)$ is the inverse of the standard $\alpha$-Fr\'echet cumulative distribution
function $\Phi_\alpha(x)=e^{-1/x^\alpha},\ x>0$. This way, in practice, under normal traffic conditions, the probability of flagging any bin in time-slot 
$t$ as a heavy hitter is no greater than $(1-p_0)$. The rate of potential false alarms, may be controlled and reduced by judiciously increasing the 
level $p_0$.  On the other hand, the presence of abnormally large bins relative to the reference distribution will be flagged if their values exceed the threshold
$T_{p_0}(m,\alpha,c)$.

To be able to use this methodology, one should estimate the key parameters $\alpha$ and $c$ appearing in formula \eqref{e:Frechet-threshold}. 
%There are many methods for estimating the parameters of the so-called extreme value distributions arising in the limit of i.i.d.\ maxima (see e.g.\ \
%\cite{embrechts:kluppelberg:mikosch:1997}).  Many of these methods involve, however, the choice of tuning parameters and/or visual inspection or 
%likelihood optimization, which is prohibitive in our context where the parameters need to be estimated on-the-fly multiple times per minute. 
The recently proposed max-spectrum method in \cite{stoev:michailidis:taqqu:2011} is particularly well-suited to this task.  It is easy to tune, 
robust to outliers, computationally efficient, and it provides estimates of both the scale parameter $c$ and the tail exponent $\alpha$.
This methodology is summarized in the formal algorithm (Algorithm~\ref{algo:heavy-hitters}).

\begin{rem} The hash-array is obtained from the {PF\_RING}-based methodology at the time scale of one array per several seconds.  For the
traffic conditions in the Merit Network (e.g.\ rates of 10Gbps), we found that time-windows of $10$ seconds provide sufficiently well-populated 
bins that lend themselves to reasonable heavy-hitter detection.  In this setting, we output estimates of heavy hitters every 10 seconds. For greater traffic rates, 
hash-binned arrays are populated faster and our methodology can be applied at an even shorter, sub-second time-scale.
\end{rem}

\begin{rem} Proposition \ref{p:Frechet-limit} is an asymptotic result. 
In our experiments with real traffic data, we found the 
approximation based on the Fr\'echet distribution to be reasonably accurate for $m$ as low as $128$ and durations about $10$ seconds.  
\end{rem}

\begin{minipage}[t]{.5\textwidth}
\vspace{-20pt}
\hspace{-30pt}
\begin{algorithm}[H]

 \caption{ \label{algo:heavy-hitters}Fr\'echet method}
 \footnotesize
 \begin{algorithmic}[1]
  \renewcommand{\algorithmicrequire}{\textbf{Input:}}
  \renewcommand{\algorithmicensure}{\textbf{Output:}}
 \REQUIRE Stream of hash-arrays $X_t = \{X_t(i)\}_{i=1}^m$; \\ probability level $p_0\in (0,1)$; \\
  smoothing coefficient $\lambda \in (0,1)$.
 \ENSURE  Stream of significant heavy-hitter bins ${\cal H}_t \subset\{1,\ldots,m\}$ and their counts $k_t = |{\cal H}_t|$.

\FOR {each stream item $X_t$}
 \STATE Estimate the tail exponent $\hat \alpha := \alpha(X_t)$ and scale coefficient $\hat c := c(X_t)$ from 
  the sample $X_t = \{X_t(i)\}_{i=1}^m$ based on 
 the {\em max-spectrum}.
  \IF {($t=1$)}
    \STATE Set $\alpha_t := \hat\alpha$ and $c_t:= \hat c$
   \ELSE  
   \STATE Perform EWMA smoothing:
  $
   \alpha_t := \lambda \hat \alpha + (1-\lambda) \alpha_{t-1}
  $ and \\
  $
    c_t:= \lambda \hat c + (1-\lambda) c_{t-1}.
  $
  \ENDIF
  \STATE Compute the significance threshold $T_t:= T_{p_0}(m,\alpha_t,c_t)$ using \eqref{e:Frechet-threshold}.
 
  \STATE Estimate the set of heavy hitter \\ bins ${\cal H}_t$ at window $t$ as 
 $
  {\cal H}_t:= {\Big\{} i\in \{1,\ldots,m\} \, :\, X_t(i) \ge T_t{\Big\}}.
 $ 
  \RETURN ${\cal H}_t$ and $k_t:= |{\cal H}_t|$.
  \ENDFOR
\end{algorithmic}

 \end{algorithm}
    \end{minipage}%
    \begin{minipage}[t]{0.5\textwidth}
    \vspace{-20pt}
\begin{algorithm}[H]
 \caption{\label{algo:topk-volume}Relative volume}
 \footnotesize
 \begin{algorithmic}[1]
  \renewcommand{\algorithmicrequire}{\textbf{Input:}}
  \renewcommand{\algorithmicensure}{\textbf{Output:}}
 \REQUIRE Stream of hash-arrays $X_t = \{X_t(i)\}_{i=1}^m$; probability level $p_0\in (0,1)$; 
 candidate value $k \in \{1,\ldots,m\}$ (preferably $\ll m$); smoothing parameter $\lambda\in(0,1)$.
 \ENSURE  Binary stream of alarm-flags $f_t\in \{0,1\}$. 

\FOR {each stream item $X_t$}
 \STATE Estimate the tail exponent $\hat \alpha := \alpha(X_t)$ from the sample $X_t = \{X_t(i)\}_{i=1}^m$.
 
  \IF {($t=1$)}
    \STATE Set $\alpha_t := \hat\alpha$ 
   \ELSE  
   \STATE Perform EWMA smoothing:
  $
   \alpha_t := \lambda \hat \alpha + (1-\lambda) \alpha_{t-1}.
  $
  \ENDIF
  \STATE Compute the relative volume of of the top--$k$ bins $V_t(k)$ as in \eqref{e:Vt}.

  \STATE Using Monte Carlo simulations, compute numerically the significance threshold 
  $q_t = q_t(p_0;k,\alpha_t,m)$,
 such that 
 $$
  \P(W_{\alpha_t}(k,m) \le q_t) \approx p_0.  
 $$ 
   
 \RETURN  $f_t:=\mathbb I \{ V_t(k) > q_t\}$, i.e., flag $V_t(k)$ as significantly large 
 (at level $p_0$) if $V_t(k) > q_t.$
 \ENDFOR
\end{algorithmic} 
 \end{algorithm}
 \vspace{20pt}
 \end{minipage}

 \subsection{Detection via Relative Volume}
 \label{sec:detectvol}

Alternatively, one can detect high-impact events by monitoring the volume
of the top-hitters relative to the total traffic. As before, 
suppose that $X_t = \{X_t(i)\}_{i=1}^m$ is a hash-binned array of traffic volume (in bytes)
computed over a given time window.  

Sort the bins in decreasing order, so that
$
X_t(i_1) \ge \cdots \ge X_t(i_k) \ge \cdots \ge X_t(i_m)\ge 0.
$
Fix a $k\in \{1,\ldots,m\}$ and consider the relative volume of traffic contributed by the top-k bins:
\begin{equation}\label{e:Vt}
V_t(k) := { \sum_{j=1}^k X_t(i_j) \over \sum_{j=1}^m X_t(i_j)}.
\end{equation}
Note that the indices of top-k bins can change from one time-window to the next. 

We aim to identify when $V_t(k)$ is `significantly' large.  For example, if $k=1$,
one would like to know if the top bin suddenly carries a very large proportion of the traffic relative to the rest.  This could
indicate an anomaly in the network.  As in the previous section, we will measure significance relative to a baseline probability model,
which is dynamically estimated from the data. The ubiquitous heavy-tailed nature of the $X_t(i)$'s will play a key role.

% We shall need the
%following fundamental `R\'enyi representation' result for the joint distribution of the {\em order statistics}
%(see, e.g., p.\ 189 in \cite{embrechts:kluppelberg:mikosch:1997}).

%\begin{thm}[R\'enyi representation] \label{thm:Renyi}
%Let $U(1),\ldots, U(m)$ be independent and identically distributed Uniform$(0,1)$ random variables. 
%Consider the sorted sample (order statistics) $U(i_1;m) \le \cdots \le U(i_k;m) \le \cdots \le U(i_m;m)$. Then, we have
%the following stochastic representation:
%$$
%{\Big(}U(i_1;m), \cdots, U(i_k;m), \cdots, U(i_m;m){\Big)} 
% \stackrel{d}{=} {\Big(} \frac{\Gamma_{1}}{\Gamma_{m+1}}, \cdots, \frac{\Gamma_k}{\Gamma_{m+1}}, \cdots, \frac{\Gamma_m}{\Gamma_{m+1}} {\Big)},
%$$
%where $\stackrel{d}{=}$ means equality in distribution and $\Gamma_i = E_1+\cdots+ E_i,\ i=1,\ldots,m+1$ are Gamma$(i,1)$-distributed random
%variables, represented as cumulative sums of a fixed set of independent standard Exponential random variables.
%\end{thm}

Let now $X(i),\ i=1,\ldots,m$ be i.i.d.\ non-negative random variables representing a generic hash-binned traffic array.  As argued in the previous
section, in a wide range of regimes, the distribution of the $X(i)$'s is heavy tailed, and they may be assumed independent because of the 
pseudo-randomization due to hashing. Thus, as in \eqref{e:Pareto-tail}, we shall assume that 
$
%begin{equation}
 %\label{e:Pareto-tail}
 \overline F(x) \equiv 1- F(x) =\P(X(1) > x) \sim c/x^{\alpha},
$%\end{equation}
~for some $c>0$ and $\alpha>0$.  It is well-known that if the distribution function $F$ is continuous, then $U(i):=\overline F(X(i)),\ i=1,\ldots,m$ are i.i.d.\ Uniform$(0,1)$. 
Therefore, the R\'enyi representation for the joint distribution of the {\em order statistics}
(p.\ 189 in \cite{embrechts:kluppelberg:mikosch:1997}), implies 
%$
%{\Big(} \overline F(X(i_1)),\cdots,  \overline F(X(i_k)), \cdots,  \overline F(X(i_m)) {\Big)} \stackrel{d}{=} {\Big(} \frac{\Gamma_{1}}
%{\Gamma_{m+1}}, \cdots, \frac{\Gamma_k}{\Gamma_{m+1}}, \cdots, \frac{\Gamma_m}{\Gamma_{m+1}} {\Big)}.
%$
%By applying the inverse function $\overline F^{-1}$ to all components of the above relation, we obtain
\begin{equation}\label{e:X-order-stat}
{\Big(} X(i_k) {\Big)}_{k=1}^m \stackrel{d}{=} {\Big(} \overline F^{-1} {\Big(}  \frac{\Gamma_k}{\Gamma_{m+1}} {\Big)} {\Big)}_{k=1}^m.
\end{equation}
This yields the following result about the distribution of the relative volume.
\begin{prop} \label{p:order-stat} 

{\bf (i)} Under the above assumptions, we have
\begin{equation}\label{e:p:order-stat-i} 
\{ V(k;m),\ k=1,\ldots,m\} \stackrel{d}{=}  {\Big\{} { \sum_{j=1}^k \overline F^{-1}(\Gamma_{j}/\Gamma_{m+1}) \over 
\sum_{j=1}^m \overline F^{-1}(\Gamma_{j}/\Gamma_{m+1})},\ k=1,\ldots,m{\Big\}}.
\end{equation}

{\bf (ii)} Under \eqref{e:Pareto-tail}, for fixed $1\le k<\ell$, we have, as $m\to\infty,$
\begin{equation}\label{e:p:order-stat-ii} 
\frac{V(k;m)}{V(\ell;m)} \stackrel{d}{\longrightarrow} W_\alpha(k,\ell) := {\sum_{j=1}^k \Gamma_j^{-1/\alpha}  \over \sum_{j=1}^\ell \Gamma_j^{-1/\alpha}}.
\end{equation}
%{\bf (iii)} Moreover, for $\overline F(x) = c/x^{\alpha}, x\ge c^{1/\alpha}$, we obtain
%\begin{equation}\label{e:p:order-stat-ii} 
%V(k;m) \stackrel{d}{\approx} W_\alpha(k,m):= {\sum_{j=1}^k \Gamma_j^{-1/\alpha}  \over \sum_{j=1}^m \Gamma_j^{-1/\alpha}}.
%\end{equation}
\end{prop}
The proof is given in the Appendix. 
Recall that our goal is to test whether $V(k;m)$ is significantly large.  The asymptotic result in \eqref{e:p:order-stat-ii} suggests that the distribution of the statistic $W_\alpha(k,\ell)$
can be used as a baseline model. Note however that it quantifies the magnitude of $V(k,m)$ relative to $V(\ell,m)$ for some fixed $\ell$.  In practice, in the context of
network traffic hash-binned arrays we studied, it turns out that $V(\ell,m) \approx 1$ for moderately large values of $\ell$.  Therefore, the denominator in the left-hand side of
 \eqref{e:p:order-stat-ii} can be taken as $1$. Further, to be slightly conservative, one can take $\ell = m$.  We therefore obtain the distributional approximation
$
V(k;m) \stackrel{d}{\approx} W_\alpha(k,m):= {\sum_{j=1}^k \Gamma_j^{-1/\alpha}  \over \sum_{j=1}^m \Gamma_j^{-1/\alpha}}.
$ 
Note that this approximation is in fact valid exactly, if $\overline F(x) = c/x^{\alpha},\ x\ge c^{1/\alpha}$, i.e.\ under the Pareto model. This discussion leads to 
Algorithm~\ref{algo:topk-volume}.

\begin{rem} In  scenarios where the Pareto approximation is not as accurate,
one can  adapt the above algorithm by considering $\ell<m$ and 
test the contribution of ratios of volumes $V(k;m)/V(\ell;m)$, relative to 
the baseline distribution of $W_\alpha(k,\ell)$. As indicated above, for simplicity, 
and to be slightly conservative in practice, we use $\ell =m$, which worked
rather well.% in our context.
\end{rem}

The significance threshold $q_t$ in Algorithm \ref{algo:topk-volume} may 
fluctuate substantially in time, since the tail exponent $\alpha_t$ does (see, e.g.\ Figure 5 in~\cite{kallitsis15amon}).
This natural {\em adaptivity} property allows us to dynamically calibrate to the changing
statistical properties of the stream. It is important, however, to be also {\em robust} to 
sudden changes of regime due to the onset of anomalies, i.e., we should not adapt to the anomalies we
are trying to detect.  Such robustness can be achieved and tuned by the smoothing parameter 
$\lambda$.  The smaller the value of $\lambda$, the closer the $\alpha_t$ to past values
$\alpha_s$,\ $s\le t$. Some degree of smoothing can also improve   estimation accuracy
through {\em borrowing strength} from the past.  In practice, we found that $\lambda \approx 0.5$ 
works well in our conditions.

If the type of anomalies considered {\em persist} over several windows of time $\Delta$, one can substantially decrease the
false alarm rate by considering {\em control charts}.  This leads to a slight modification of
Algorithm \ref{algo:topk-volume}. Following \cite{lambert:liu:2006}, one can consider
the p-values,
$
p_t:= \P(V_t(k) > W_{\alpha_t}(k,m)),
$
and then apply an EWMA on the {\em z-scores}:
$
z_t := \lambda_p \Phi^{-1}(1-p_t) + (1-\lambda_p) z_{t-1},
$
for another weight $\lambda_p \in (0,1)$. Then, under baseline conditions, $z_t$ follows the Normal distribution with
zero mean and variance $\sigma_z^2 =\lambda_p/(2-\lambda_p)$. Thus, classical process control methodology suggests to raise
an alarm if $z_t/\sigma_z >L$, for a given level parameter $L>0$~\cite{Lucas:1990:EWM:84840.84843}. 
The pair of parameters $(\lambda_p,L)$ can be tuned so as to ensure detection of persisting anomalies, while minimizing false alarms.
Section \ref{sec:peva} includes studious sensitivity analyses of these calibrations controls.

%\vspace{-20pt}
% \subsection{Detection via Relative Volume}
% \label{sec:detectvol}
 %\input{volume.tex}

\vspace{-15pt}
 \subsection{Community Detection}
 \label{sec:detectcomm}
 
Consider now the two-dimensional matrix $X_t = \{X_t(i,j)\}_{i,j=1}^m$ of updates, obtained for a certain time $t$.  
The technique proposed next aims at detecting changes in 
the {\em community structure} of the network flows. To this end, focus on the top $N$ bins of $X_t(i,j),\ i,j=1,\ldots,m$,
which represent an aggregate summary of the top origin-destination flows in the network.
 
Let $A_t = (a_t(i,j))_{m\times m}$ be a binary matrix, such that $a_t(i,j)=1$ if and only if bin $(i,j)$ belongs to the set of top $N$ items in the array $X_t$.
One can view $A_t$ as an adjacency matrix of an oriented graph $G_t$, which is a type of a histogram of the underlying (rather sparse) graph of flows from
a given sIP to a dIP that are active over the time-window of interest.  Changes in the connectivity of $G_t$ indicate changes in the community structure of
the traffic flows. For example, in the event of a DDoS or other distributed attacks, a given destination IP $\omega_0$ is flooded with substantial amount of  traffic from multiple 
source IPs.  If a large number of sources are involved, then this will likely result in a horizontal strip of relatively large values in 
the two-dimensional hash-binned array.  The location of the strip will be $i_0:= h(\omega_0)$---the index of the bin where
the target destination IP $\omega_0$ is hashed
(see, e.g.\ Figures~\ref{fig:tor_case} and~\ref{fig:vis_tools} for visualizing the `Tor', `SSH-scanning' and `SSDP' attack events).

One way to formally and automatically  detect such features is to focus on the graph with adjacency matrix $A_t$.  In this event, the matrix $A_t$ will 
have a relatively larger number of $1$s in the $i_0$th row and, correspondingly, the in-degree of node $i_0$ will be large. We propose a statistical 
method for quickly identifying {\em significant} peaks in the in-degrees (or out-degrees). This method, combined with the information from the Boyer-Moore MJRTY instances 
associated with the bins involved can lead to an almost instantaneous identification of possible targets as well as (potential) culprits of malicious activities.

Focus on ingress connectivity, i.e., let
$
I_t(i):= \sum_{j=1}^m a_t(i,j),\ \ i=1,\ldots,m
$ 
be the in-degree associated with node $i$ for the oriented graph $G_t$.  Our goal is to flag statistically significant peaks of $I_t(i)$.  As argued in Sections \ref{sec:detecthh}
and \ref{sec:detectvol}, hashing ensures randomization and hence $I_t(i),\ i=1,\ldots,m$ can be reasonably assumed to be independent. In contrast with the previous sections,
however, the distribution of the counts $I_t(i)$ are no longer heavy-tailed but rather well-approximated by a Normal distribution.  For a fixed $i$, thanks to the randomization 
induced by hashing, one can view $a_t(i,j)$'s as independent in $j$. Hence, for relatively large $m$, as well as $N$ and ultimately traffic rate, the CLT ensures that centered and normalized integer
counts $I_t(i)$ can be modeled by the Normal distribution. Indeed, Figure \ref{fig:src_series_conn_dist} (right panel, top-right plot) shows Normal quantile-quantile plots of $I_t(i),\ t=1,\ldots,T$
for $5$ typical (non-anomalous) bins $i$. The linearity in the plots indicates agreement with the Normal distribution. The heatmap therein (top-left) shows the entire array $(I_t(i))_{m\times T}$
of in-degrees computed over $10$-second time windows over the duration of 1 hour.  We focused on the top $N=3000$ flows.  The bottom plots in this figure show the
QQ-plots corresponding to anomalous bins with high in-degree corresponding to the higher intensity lines in the top-left plot.

%\begin{figure}[ht!]
%\includegraphics[width=0.6\textwidth]{./figures/Connectivity_Distributions.eps}
%\caption{Merit Network 16:00-17:00, Aug 1, 2015 -- the `Tor' event in Section \ref{subsec:tor}.
%{\em Top-left:} Ingress connectivity for the top $N=3000$ hash-binned flows per 10-second windows over 1-hour. {\em Top-right:} QQ-plots %demonstrating
%accuracy of the Normal approximation of typical in-degree distributions. {\em Bottom plots}: QQ-plots for anomalous bins. }
%\label{fig:connectivity-distributions}
%\end{figure}

Given the above discussion, in the baseline regime, we shall assume that
$I_t(i),\ i=1,\ldots,m$ are independent ${\cal N}(\mu_t,\sigma_t^2)$. Then for 
$D_t:= \max_{i=1,\ldots,m} I_t(i),$
by the independence of the $I_t(i)$'s, we obtain 
$
\P(D_t \le x) = \Phi{\Big(} {x-\mu_t \over \sigma_t }{\Big)}^m,
$
where $\Phi$ is the standard normal CDF.  Fix a probability level $p_0$ (e.g.\ $0.99$), and consider the {\em significance threshold}
$
 u_t(p_0)\equiv u(p_0,m,\mu_t,\sigma_t) := \mu_t + \sigma_t\times p_0^{1/m}.
$ 
Thus, in the baseline regime, all in-degrees $I_t(i),\ i=1,\ldots,m$ lie below $u_t(p_0)$ with probability $p_0$.  As in Section \ref{sec:detecthh}, we
shall flag all bins $i$, for which $I_t(i)$ exceeds $u_t(p_0)$ as {\em anomalous}.  The detection algorithm is analogous to Algorithm \ref{algo:heavy-hitters}, except
that now one should estimate the parameters $\mu_t$ and $\sigma_t$. This can be similarly done using an EWMA of the empirical means and standard deviations
of the samples $I_t(i),\ i=1,\ldots,m$.  We omit the details to avoid repetition.

This method is illustrated
in Section~\ref{sec:peva}, where it is successfully employed in mining 
seemingly harmless  events characterized by high node-connectivity (e.g., the `SSDP' and `Tor' cases). 
These events are harder to detect via the methods described in 
Sections \ref{sec:detecthh} and \ref{sec:detectvol}.

\begin{rem} Observe that the access to various cloud services and resources can have similar characteristics, where multiple source IPs communicate 
with a single destination IP (server).  Such servers, however, are typically well-known and can be {\em a priori} filtered out.  
An alternative application of this methodology is to track the up-surge of users to a particular service, such as Twitter, Facebook or Google, for example.  
Such up-surges, not necessarily due to malicious activity, may be of interest to network engineers or researchers.
\end{rem}

 %\section{Applications}
 %
 %  \subsection{Online: interactive monitoring}
 %   \textcolor{red}{Good to mention as a future feature, but we don't have an interactive software right now. Good idea for proposal}
 %  
 %  \subsection{Offline: analysis and forensics}  
 %  \textcolor{red}{Besides the online creation of data bricks (or 1D bricks) all the rest is near-online.}
\vspace{-15pt}
\section{Performance Evaluation}
\label{sec:peva}
\vspace{-7pt}
%In this section, we evaluate the performance of the proposed system.
%Our assessment is based on real-world data collected at Merit.
\subsection{Software performance}

The excellent performance of PF\_RING is well documented~\cite{Fusco:2010:HSN:1879141.1879169, Gallenmuller:2015:CFH:2772722.2772729};
this section focuses  instead on our monitoring application. We perform 
measurements in situations where {AMON} is deployed in the field,
and under heavy stress-testing with a traffic generator appliance. 

Figure~\ref{fig:pfring_chicago} (right) illustrates our software capabilities when monitoring
traffic at Merit's main peering point in Chicago. Our setting involves
a passive monitor (i.e., packet tap) receiving traffic from four
SPAN 10GE ports. The mirrored traffic includes both ingress and egress
network traffic, and a 5-day snapshot of aggregate volume is shown
in~Figure~\ref{fig:pfring_chicago}. Note that traffic rates are well above 20Gbps;
however,  AMON monitoring traffic \emph{from all four
10GE ports simultaneously}, experienced minimal packet drops (below 1.5\%). Further,
the amount of physical memory required by our application was only around 40MB,
something expected from the low space complexity of the Boyer-Moore algorithm~\cite{kallitsis14hashing}.

To shed more light into this, we undertook
performance tests using a traffic generator with 40 byte payload packets 
(i.e., sending at the minimum frame of 64 bytes).
At wire-speed of 10Gbps we measured throughput that exceeded 12 Mpps (million packets per second).
This corresponds to a drop rate of $18\%$; testing with payloads of size 96 and above showed
zero loss at wire speeds. We conjecture that the bottleneck seems to be the buffer size of the NIC card
we used, and not PF\_RING. In particular, the Intel card we tested has buffers of size 4096,
and hence packet drops seem to be inevitable at these rates.%\footnote{This is also confirmed from the product brief that claims wire-rate throughput with packets $>$ 64 bytes.}.
We are in the process of conducting tests on cards with larger buffers in order to verify our hypothesis.

\vspace{-15pt}
\subsection{Identification accuracy}

% flow-stat < /data/netflow/current/WSUb/2015/2015-07/2015-07-22/ft-v05.2015-07-22.120000-0400
%Total Flows                     : 44029680
%Total Octets                    : 300397729549
%Total Packets                   : 449847167

% flow-stat < /data/mgkallit_scratch/jsac15_data/WSUe_SSDP_ft-v05.2015-12-09.110000-0500
%Total Flows                     : 92328600
%Total Octets                    : 447104219945
%Total Packets                   : 581468447

Next, we demonstrate the identification accuracy of MJRTY Boyer-Moore;
we perform comparisons against  Combinatorial Group Testing (CGT)~\cite{Cormode1061325}. 
We utilize an hour-long NetFlow dataset, collected at Merit,
with 92 million flows and an aggregate volume of 447 GBytes and around 580 million packets. 
Both methods are evaluated against the ground truth (i.e., exact recovery of top-$K$ hitters).
All methods report their answers every 100,000 NetFlow records; Table~\ref{tab:identcomp}
illustrates the average proportion of 
identified heavy hitters among the top-$K$ and the standard error (in parenthesis). 
The chosen data contain a low-volume DDoS attack attributed to
the Simple Service Discovery Protocol (SSDP); see Figure~\ref{fig:vis_tools} (left).

 The CGT method~\cite{Cormode1061325} is a probabilistic technique,  based on the ideas
 of `group testing'. It aims at finding the elements whose volume is
 at least $1/(k+1)$ of the total; this is a relaxed version of the top-$K$ hitters problem.
 The authors provide performance guarantees with respect to accuracy, space and time.
 It is suited for high-speed streaming data; indeed, besides its offline evaluation on accuracy, 
 \emph{we have implemented the method in the AMON framework
 and verified its time and space efficiency.} Its online realization demonstrated results similar to Figure~\ref{fig:pfring_chicago}. 
 For the results of Table~\ref{tab:identcomp} we sought the top source IPs per interval.
 The tuning parameters for CGT include the hash-table size $W$, and the number of
 groups $T$ (in all experiments, $T=2$; increasing $T$ improves accuracy
 but worsens the efficiency on real-data). The granularity unit $b$ (see~\cite{Cormode1061325}, Sec. 3.3)
 is set to $b=8$ for better efficiency, at the expense of space in memory. 

The MJRTY BM is regulated with the number of sub-streams, $m$. Note that, regarding space utilization,
hashing with size $m=1024$ corresponds to the CGT case $W=1024$. In all cases,
we used $W\ge 2k$, per Lemma 3.3 in~\cite{Cormode1061325}. The
results of Table~\ref{tab:identcomp} showcase that MJRTY BM is highly
accurate in finding the most frequent elements of the stream. It also often outperforms
its competitor.  CGT's performance can increase with elevated
values of $W$ and $T$ at expense of space and, most importantly, time. 
However, MJRTY BM can increase its accuracy too by stretching $m$.
Finally, we note that CGT, being a probabilistic algorithm,
may output IP elements that are not present in the stream (due to hash collisions).
Conversely, MJRTY BM is not susceptible to this.

%As described above, the Boyer-Moore algorithm guarantees returning the majority element on a given thinned stream, when there is one.
%We collected these fractions
%during the 5-day experiment with the Chicago traffic (Figure~\ref{fig:pfring_chicago}),
%and on average at least $85.41\%$ of all sub-streams (with $0.11\%$ standard deviation)
%contained a majority element.

%\begin{figure}[t!]
%        \centering
%		\includegraphics[width=0.45\textwidth]{./figures/zc_mon_performance.pdf}
%\vspace{-20pt}
%\caption{\footnotesize{Software performance (Chicago site); rates exceeded 20Gbps, but minute drop rates recorded. }}
%\label{fig:pfring_chicago}
%\vspace{-10pt}
%\end{figure}

%\begin{figure*}[t!]
%        \centering
%        \begin{subfigure}[b]{0.45\textwidth}
%               \includegraphics[width=.85\textwidth]{./figures/bm_accuracy_bytes.pdf}
%        \end{subfigure}
%        \begin{subfigure}[b]{0.45\textwidth}
%                 \includegraphics[width=.85\textwidth]{./figures/bm_accuracy_packets.pdf}
%        \end{subfigure}
%        \vspace{-15pt}
%	\caption{\footnotesize{Boyer-Moore accuracy. {\em Left:} Hitters based on bytes. {\em Right:} Packets.}}
%\label{fig:bm_eval}
%\vspace{-25pt}
%\end{figure*}

\begin{table}[t]
\centering
\caption{Identification; comparison with Combinatorial Group Testing (CGT)~\cite{Cormode1061325}.}
\vspace{-10pt}
\label{tab:identcomp}
\begin{tabular}{l|cc|ccc}
Top-K hitters   &  BM (m=512)   &  BM (m=1024)  & CGT (k=500,W=1024)      & CGT (k=1000,W=2048)       & CGT (k=2000,W=4096)  \\\hline
K=10 (packets) & $0.99_{(0.04)}$        & $0.99_{(0.03)}$            & $0.91_{(0.10)}$     & $0.91_{(0.10)}$     &  $0.89_{(0.09)}$ \\
K=10 (bytes)     & $0.99_{(0.03)}$        & $0.99_{(0.03)}$            &  $0.79_{(0.14)}$   & $0.85_{(0.12)}$     & $0.86_{(0.11)}$ \\

K=50 (packets)  & $0.94_{(0.04)}$       & $0.98_{(0.02)}$            &  $0.90_{(0.07)}$   & $0.90_{(0.05)}$     & $0.90_{(0.04)}$ \\
K=50 (bytes)      & $0.96_{(0.03)}$       & $0.98_{(0.02)}$            &  $0.80_{(0.11)}$   & $0.89_{(0.06)}$     & $0.90_{(0.05)}$ \\

K=100 (packets)  & $0.85_{(0.05)}$      & $0.94_{(0.03)}$           &  $0.60_{(0.11)}$   & $0.89_{(0.05)}$    & $0.90_{(0.03)}$ \\
K=100 (bytes)     & $0.90_{(0.03)}$       & $0.96_{(0.02)}$           &   $0.58_{(0.09)}$  & $0.86_{(0.09)}$    & $0.92_{(0.04)}$ \\

K=200 (packets)  & $0.71_{(0.05)}$      & $0.87_{(0.03)}$           &  $0.30_{(0.06)}$   & $0.76_{(0.12)}$    & $0.91_{(0.03)}$ \\
K=200 (bytes)     & $0.77_{(0.04)}$       & $0.90_{(0.02)}$           &  $0.29_{(0.05)}$   & $0.48_{(0.08)}$    & $0.78_{(0.13)}$
\end{tabular}
\vspace{-30pt}
\end{table}

\vspace{-15pt}
\subsection{Detection accuracy}
\begin{figure*}[t!]
        \centering
        \begin{subfigure}[b]{0.24\textwidth}
               \includegraphics[width=1\textwidth]{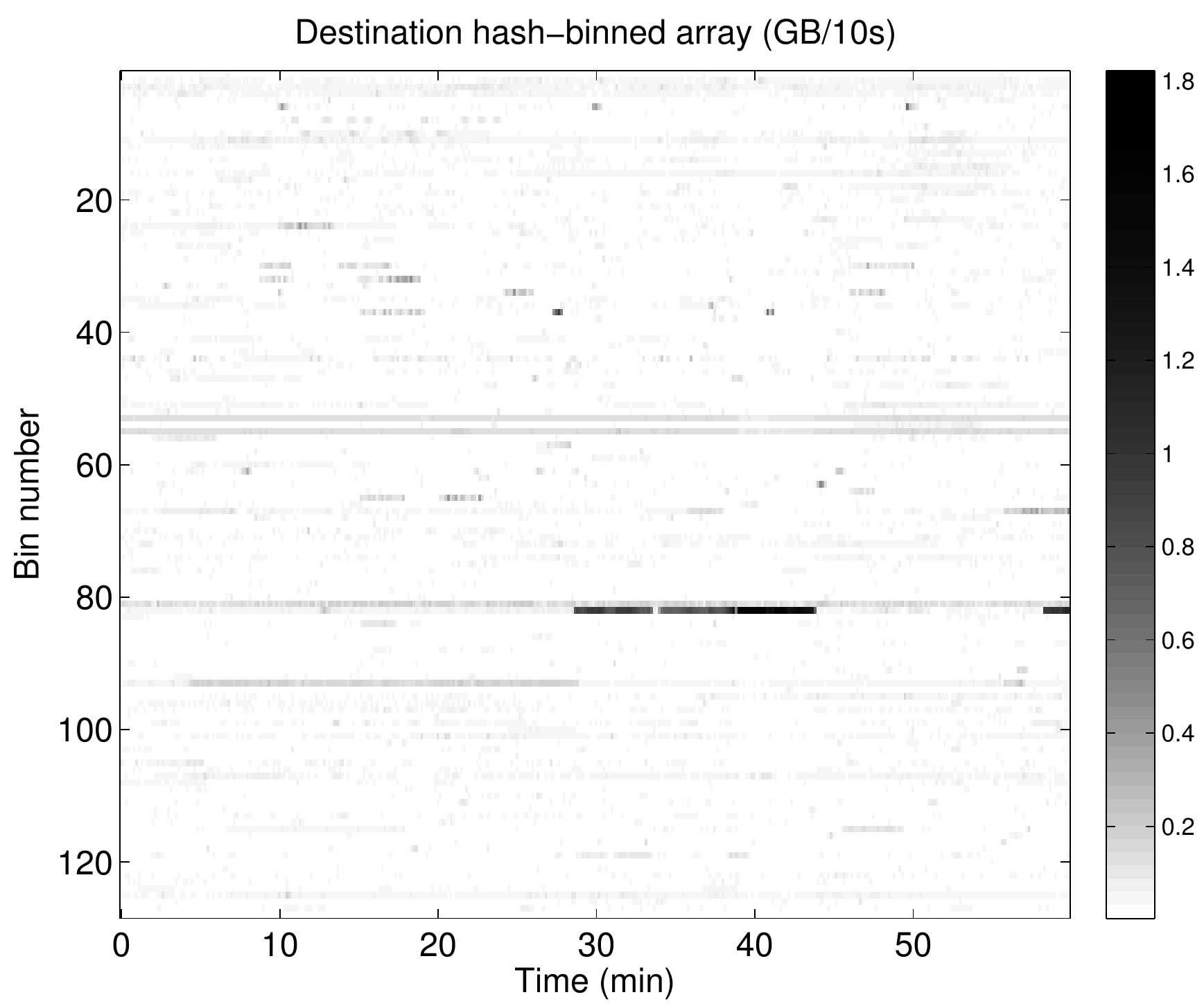}
        \end{subfigure}
        \begin{subfigure}[b]{0.24\textwidth}
                 \includegraphics[width=1\textwidth]{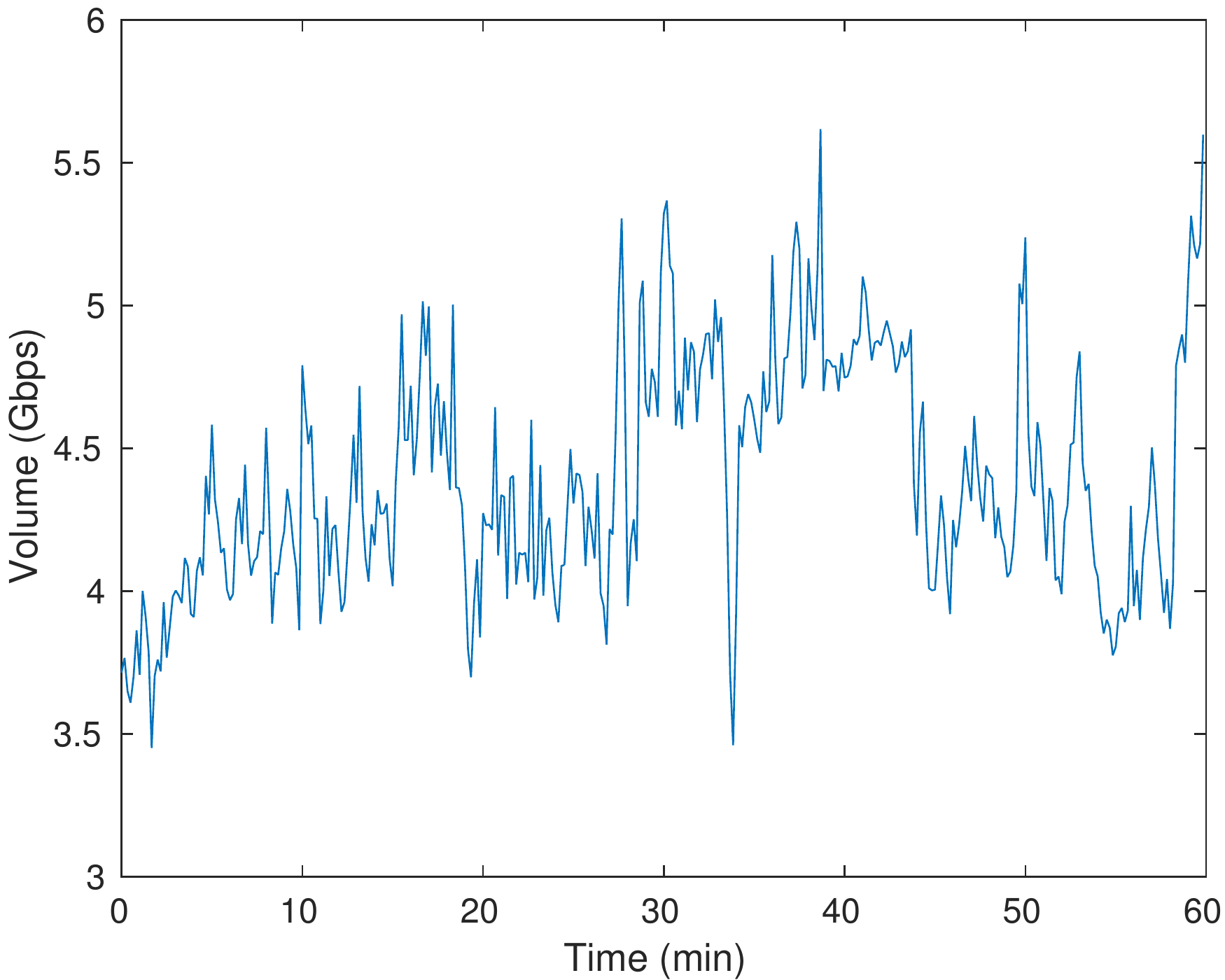}
        \end{subfigure}
        \begin{subfigure}[b]{0.24\textwidth}
                 \includegraphics[width=1\textwidth]{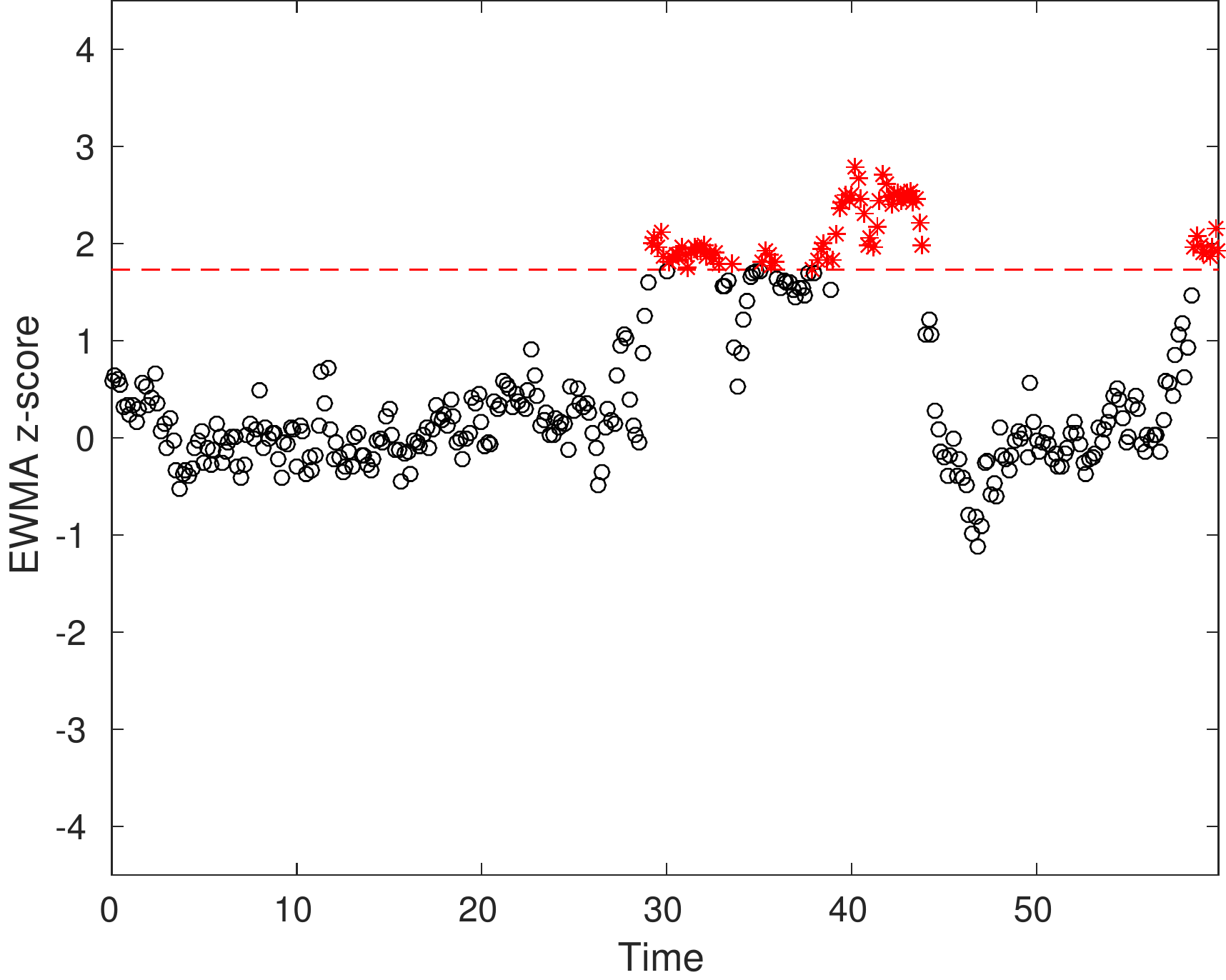}
        \end{subfigure}
        \begin{subfigure}[b]{0.24\textwidth}
                 \includegraphics[width=1\textwidth]{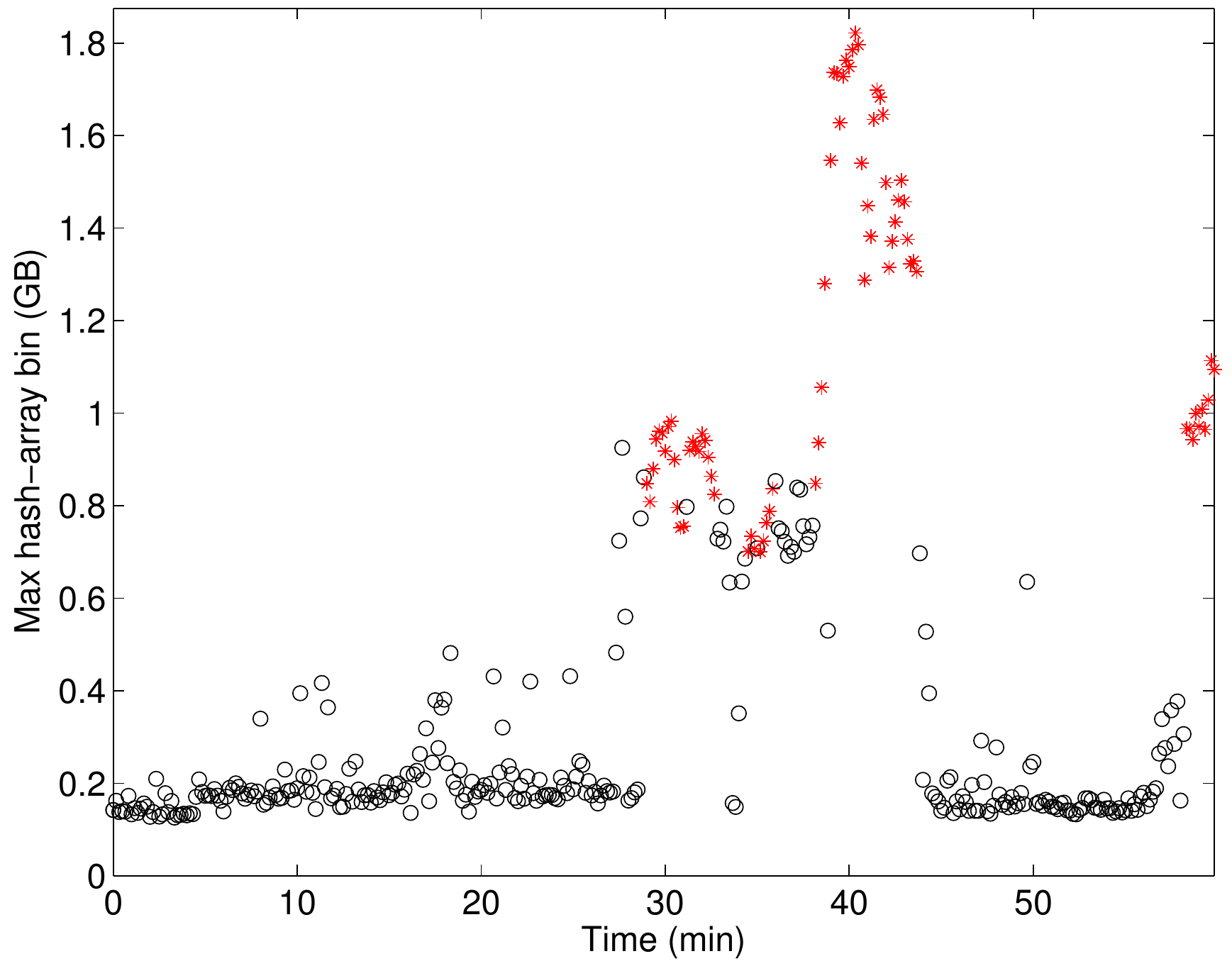}
        \end{subfigure}
 \vspace{-10pt}
 \caption{\footnotesize{The `Library' event. 
 The left panel shows the time-series of the hashed-array
for destinations for a period of one hour. Note the dark horizontal
stripe (at bin 82) between minutes 30 and 45 and towards the end of the hour. The adjacent panel depicts the aggregate traffic volume
over the hour of interest. Observe the elevated traffic volume and note that both 
the Fr\'echet and the relative volume methods
correctly raised alerts (red) during the malicious activity period (rightmost figures).}}
\label{fig:library_case}
\vspace{-25pt}
\end{figure*}

\vspace{-5pt}
We shed light into the detection accuracy of our methods by considering real-world DDoS case studies
as well as synthetic attacks injected on real data. 
The studied attacks were recorded at Merit's NetFlow collector.
The first event, labeled as `Library' case study, involved heavy
UDP-based DNS and NTP traffic to an IP registered to a public library in
Michigan, and is considered a volumetric attack (see Figure~\ref{fig:library_case}). The second event, named `SSDP',
is a low-volume attack directed to another host within the network (see Figure~\ref{fig:vis_tools}, left). 

We implemented the \emph{Defeat}~\cite{Li:2006:DIN:1177080.1177099} subspace method
and juxtapose its performance against our algorithms on the two attacks. 
\emph{Defeat} checks for anomalies using principal component analysis (PCA).
A dictionary of entropies is built, over moving windows, of distributions of certain signature signals involving source and 
destination ports and IP addresses. 
In \emph{Defeat}, abnormalities are viewed as unusual distributions of these
features. %In particular, the \emph{entropy} of the empirical distribution of each feature is used to detect unusual traffic patterns. 
The \emph{Defeat} framework is well-suited for multi-dimensional data due to its sketch-based design,
and can be utilized for detection, identification and classification of attacks. However, its requirement
for construction of empirical histograms makes it less appealing (if any feasible at all) for online realization.

Table~\ref{tab:beatdefeat} tabulates our analyses. We 
report two metrics, namely \emph{precision} and \emph{recall}.
Precision depicts the fraction of alerts raised that are indeed relevant, %i.e., Tp/(Tp+Fp),
and recall captures the ratio of actual anomalies that were detected. %, that is Tp/(Tp+Fn).
The ground truth (i.e., instances that the attack was ongoing) was obtained by offline data analysis that revealed
the times when the target IPs and the corresponding protocol ports ranked among the top-10.
Again, we considered time windows of 100,000 flow records. 
Due to the fact that additional attacks 
unknown to us might be present in the data, the precision criterion should be interpreted
as a worst-case, lower bound. 
%For the `Library' case
%this translates to around 500 time bins (an hourly long interval); for `SSDP', there exist 1900  bins (over a two-hour interval).

The \emph{Defeat} method was calibrated by the significance level $\alpha$ (necessary for its monitoring statistic threshold)
and the number of `votes' raised by \emph{Defeat}'s internal detection processes. A sketch of size 484 was employed.
For our system, we ran all three detection methods and reported their results; we also
demonstrate the overall AMON accuracy by accounting the union of alerts.
As illustrated in Table~\ref{tab:beatdefeat}, both \emph{Defeat} and AMON perform remarkably well on  `Library'.
Recall that this event is a voluminous one, and hence both techniques can easily
detect it. On the other hand, the `SSDP' case is a harder one (see Figure~\ref{fig:vis_tools}). \emph{Defeat} reports true alerts for a higher time fraction,
and both methods show consistent and similar false positive rates.
The fact that \emph{Defeat} checks for more traffic features than our methods seems to be the explanation
of its higher attack discovery rate. However, we emphasize that AMON indeed rapidly uncovered the underlying event
during its period of appearance.
%for about half the times it occurred. 
Further, it is extensible and adding monitoring
features like source and destination ports into its design is straightforward.

The \emph{Defeat} method works well but is rather sophisticated and requires a {\em training period}. 
This training process is computationally intensive and has to be performed offline.  Moreover,
the construction of signal distributions and entropy calculations, required to {\em run} the PCA-based detection requires very
large memory structures, which do not scale well in real network conditions.  Making this method work in real-time requires
a formidable and independent effort. 
%Arguably, \emph{Defeat}'s detection performance is very robust. Nevertheless, we observed in first-hand
%how cumbersome might be for operators to fine tune it and `train' it with real-world data. \emph{Defeat}'s
%detection mechanism is based on finding a `normal' subspace by applying a singular value decomposition
%on data from a training period with no anomalies. Further, traffic should be \emph{stationary}
%in the selected training window. 
Further, finding a sufficiently long, anomalous-free period that satisfies the stationarity assumption might be challenging. 
%While robust PCA techniques~\cite{Candes:2011:RPC:1970392.1970395} could help alleviate partially the problem,
Its adaptability in dynamically changing traffic conditions  is another concern. In contrast, our methods
are highly adaptive to traffic trends and require no training. 
\begin{table}
\centering
\caption{Detection accuracy; comparison with \emph{Defeat}~\cite{Li:2006:DIN:1177080.1177099}.}
\label{tab:beatdefeat}
\vspace{-15pt}
\begin{subtable}{.5\textwidth}
\centering
%\begin{table}[t]
%\centering
%\footnotesize
%\caption{Comparisons.}
%\label{tab:compare_defeat}
\caption{The Library case study}
\vspace{-5pt}
\begin{tabular}{lll}
                           Method                 			 & Prec. & Recall \\\hline\hline
 \emph{Defeat} ($\alpha = 0.01$)                  		   & 0.95      & 0.94   \\
 \emph{Defeat} ($\alpha = 0.001$)                  		  & 0.80      & 0.95   \\\hline
Fr\'echet   ($p=0.95, \lambda_\alpha=0.6$)      & 0.89      & 0.22   \\
Rel. Vol.  ($\lambda_p=0.6, L=1.64$)                     & 0.80      & 0.48   \\
Connectivity ($p=0.9999$)              			   & 0.74      & 0.95   \\
\textbf{AMON (all methods)}                         	  & \textbf{0.94}      & \textbf{0.93}   \\\hline
Fr\'echet   ($p=0.85, \lambda_\alpha=0.6$)      & 0.65      & 0.41   \\
Rel. Vol. ($\lambda_p=0.6, L=1.64$)                     & 0.80      & 0.48   \\
Connectivity ($p=0.9999$)              			   & 0.74      & 0.95   \\
\textbf{AMON (all methods)}                         			   & \textbf{0.94}      & \textbf{0.93}   \\\hline
Fr\'echet  ($p=0.95, \lambda_\alpha=0.6$)      & 0.89      & 0.22   \\
Rel. Vol. ($\lambda_p=0.6, L=2$)          		   & 0.80      & 0.29   \\
Connectivity ($p=0.9999$)              			   & 0.74      & 0.95   \\
\textbf{AMON (all methods)}                         			   & \textbf{0.95}      & \textbf{0.93}   
\end{tabular}
\end{subtable}%
\begin{subtable}{.5\textwidth}
\centering
\caption{The SSDP case study}
\label{tab:ssdp}
\vspace{-5pt}
\begin{tabular}{lll}
                           Method                 			 & Prec. & Recall \\\hline\hline
 \emph{Defeat} ($\alpha = 0.001$, 9 votes)                  		   &  0.35    &  0.80    \\
 \emph{Defeat} ($\alpha = 0.001$, 10 votes)                  		  &   0.31    &  0.21    \\\hline
Fr\'echet  ($p=0.95, \lambda_\alpha=0.6$)     			 &  0.40     & 0.03   \\
Rel. Vol. ($\lambda_p=0.6, L=1.64$)                    		 & 0.47      & 0.12   \\
Connectivity ($p=0.9999$)              			   		& 0.34      & 0.45   \\
\textbf{AMON (all methods)}                         	  & \textbf{0.33}      & \textbf{0.46}   \\\hline
Fr\'echet  ($p=0.85, \lambda_\alpha=0.6$)      & 0.36      & 0.13   \\
Rel. Vol. ($\lambda_p=0.6, L=1.64$)                     & 0.47      & 0.12   \\
Connectivity ($p=0.9999$)              			   & 0.34      & 0.45   \\
\textbf{AMON (all methods)}                         			   & \textbf{0.33}      & \textbf{0.47}   \\\hline
Fr\'echet  ($p=0.95, \lambda_\alpha=0.6$)      & 0.40      & 0.03   \\
Rel. Vol. ($\lambda_p=0.6, L=2$)          		   & 0.52      & 0.06   \\
Connectivity ($p=0.9999$)              			   & 0.34      & 0.45   \\
\textbf{AMON (all methods)}                         			   & \textbf{0.34}      & \textbf{0.46}   
\end{tabular}
\end{subtable}
\vspace{-30pt}
\end{table}

To grasp insights into AMON's sensitivity to various tuning parameters we can study Table~\ref{tab:detection_hh} and
Table~\ref{tab:detection_order}.
For this evaluation, we utilized data collected 
during a seemingly ordinary period. % (we chose one hour over a holiday weekend in July). 
We randomly injected attacks of various magnitudes at 5 times; the \emph{injected traffic
volume occurs directly on the databrick matrices}. 
We first considered the scenario of many sources sending traffic to one
destination. 
In this scenario, one databrick row is `inflated'  
by the synthetic attack magnitude at 5 random instances.
The algorithm input was the destinations hash-binned arrays
(see Figure~\ref{fig:data_products}), and we allowed a grace period of 3 minutes for detection.
Each individual experiment was repeated 50 times and we report
the average performance in terms of precision, $P_d^{(1)}$, and recall, $R_d^{(1)}$; sub-script `$d$' denotes that  the algorithm
input was the destinations' signal.
We also considered the scenario of one source communicating with multiple destinations (see $P_s^{(2)}$ and $R_s^{(2)}$),
and the case of several sources to various destinations (rightmost four columns).

Table~\ref{tab:detection_hh} tabulates results for Algorithm~\ref{algo:heavy-hitters},
which is tuned by the significance level $p_0=p$ and the smoothing
parameter $\lambda=\lambda_{\alpha}$. Best performance is achieved with
$p=0.95$ and $\lambda_\alpha=0.50$. Note that $p$
may be calibrated to ease the false alarm rate. Further, observe
the connection between robustness and adaptivity
as dictated by $\lambda_\alpha$. Recall that this parameter
is used to smooth the heavy-tail exponent, $\alpha$. Big traffic spikes
translate to a heavier distribution tail and thus a low $\alpha$; this could make our 
scheme too insensitive/conservative if we do not have an adaptive
scheme that accounts for `historical' $\alpha$'s. On the other hand, high
$\alpha$ can make our scheme too sensitive (i.e., many false positives).
Table~\ref{tab:detection_order} illustrates the detection performance of a modification of Algorithm \ref{algo:topk-volume}, 
which utilizes EWMA control charts on z-scores, as explained at the end of Section \ref{sec:detectvol}.
We employ our methodology for the EWMA $(\lambda_p, L)$ pairs shown and $\lambda=\lambda_\alpha$ was fixed to $0.5$. 
For this evaluation, our synthetic attacks were persistent for 5 consecutive time slots (i.e., 50 seconds).
%The pair (0.60, 2) suggests itself based on our results.
Users can tame the alert rate by increasing the control 
limits with a higher $L$ and/or decrease further $\lambda_p$.

\begin{table*}[t]
\centering
\caption{Fr\'echet method (Algorithm \ref{algo:heavy-hitters})}
\label{tab:detection_hh}
\begin{tabular}{ccc||cc|cc|cccc}
\footnotesize
 $p$  & $\lambda_\alpha$ & {Gbps} &  $P_d^{(1)}$ &  $R_d^{(1)}$ &  $P_s^{(2)}$  & $R_s^{(2)}$ &   $P_s^{(3)}$  & $R_s^{(3)}$ &  $P_d^{(3)}$ &  $R_d^{(3)}$\vspace{3pt}\\
 \hline

     $0.95$  &   $0.50$  &   $0.50$ &    $0.74$  &  $1.00$  &  $1.00$  &  $0.96$  &    $1.00$  &  $1.00$  &  $0.74$  &  $1.00$\\
     $0.95$  &   $0.50$  &   $1.50$ &  $0.73$  &  $1.00$  &  $1.00$  &  $1.00$  &    $1.00$  &  $1.00$  &  $0.74$  &  $1.00$\\
     $0.95$  &   $0.50$  &   $2.50$ &  $0.73$  &  $1.00$  &  $1.00$  &  $0.98$  &    $1.00$  &  $1.00$  &  $0.74$  &  $1.00$\\[4pt]
     $0.95$  &   $0.60$  &   $0.50$ &  $0.72$  &  $0.93$  &  $1.00$  &  $0.61$  &    $1.00$  &  $1.00$  &  $0.85$  &  $1.00$\\
     $0.95$  &   $0.60$  &   $1.50$ &  $0.74$  &  $0.99$  &  $1.00$  &  $0.88$  &    $1.00$  &  $0.99$  &  $0.85$  &  $0.99$\\
     $0.95$  &   $0.60$  &   $2.50$ &  $0.73$  &  $1.00$  &  $1.00$  &  $0.92$  &    $1.00$  &  $1.00$  &  $0.85$  &  $1.00$\\[4pt]
     $0.99$  &   $0.50$  &   $0.50$ &  $1.00$  &  $0.73$  &  $0.74$  &  $0.22$  &    $1.00$  &  $1.00$  &  $1.00$  &  $0.99$\\
     $0.99$  &   $0.50$  &   $1.50$ &  $1.00$  &  $0.94$  &  $0.98$  &  $0.50$  &    $1.00$  &  $1.00$  &  $1.00$  &  $1.00$\\
     $0.99$  &   $0.50$  &   $2.50$ &  $1.00$  &  $0.97$  &  $1.00$  &  $0.71$  &    $1.00$  &  $0.99$  &  $1.00$  &  $0.99$\\[4pt]
     $0.99$  &   $0.60$  &   $0.50$ &  $0.76$  &  $0.24$  &  $0.36$  &  $0.09$  &    $1.00$  &  $1.00$  &  $1.00$  &  $1.00$\\
     $0.99$  &   $0.60$  &   $1.50$ &  $0.92$  &  $0.40$  &  $0.08$  &  $0.02$  &    $1.00$  &  $1.00$  &  $1.00$  &  $1.00$\\
     $0.99$  &   $0.60$  &   $2.50$ &  $1.00$  &  $0.55$  &  $0.16$  &  $0.04$  &    $1.00$  &  $1.00$  &  $1.00$  &  $0.99$
\end{tabular}
\end{table*}

\begin{table}[t]
\centering
\caption{Relative volume method (Algorithm~\ref{algo:topk-volume})}
\label{tab:detection_order}
\begin{tabular}{ccc||cc|cc|cccc}
L    & $\lambda_p$ & Gbps & $P_d^{(1)}$ &  $R_d^{(1)}$ &  $P_s^{(2)}$  & $R_s^{(2)}$ &   $P_s^{(3)}$  & $R_s^{(3)}$ &  $P_d^{(3)}$ &  $R_d^{(3)}$  \\\hline
2.00 & 0.50        & 0.50 & 0.45 & 0.98 & 0.61 & 0.97 & 0.49 & 0.99 & 0.35 & 0.99 \\
2.00 & 0.50        & 1.50 & 0.45 & 0.99 & 0.62 & 0.96 & 0.41 & 0.99 & 0.33 & 1.00 \\
2.00 & 0.50        & 2.50 & 0.45 & 0.98 & 0.64 & 0.98 & 0.41 & 0.99 & 0.34 & 1.00 \\[4pt]
2.00 & 0.60        & 0.50 & 0.61 & 0.98 & 0.80 & 0.96 & 0.74 & 0.99 & 0.45 & 1.00 \\
2.00 & 0.60        & 1.50 & 0.62 & 0.97 & 0.81 & 0.99 & 0.50 & 0.99 & 0.41 & 0.99 \\
2.00 & 0.60        & 2.50 & 0.60 & 0.98 & 0.81 & 0.97 & 0.48 & 0.99 & 0.40 & 0.99 \\[4pt]
3.00 & 0.50        & 0.50 & 1.00 & 0.62 & 0.76 & 0.25 & 0.95 & 0.99 & 0.78 & 0.99 \\
3.00 & 0.50        & 1.50 & 0.98 & 0.74 & 0.62 & 0.16 & 0.73 & 0.98 & 0.60 & 0.98 \\
3.00 & 0.50        & 2.50 & 1.00 & 0.81 & 0.90 & 0.36 & 0.58 & 0.98 & 0.55 & 0.99 \\[4pt]
3.00 & 0.60        & 0.50 & 0.96 & 0.42 & 0.50 & 0.13 & 0.99 & 0.98 & 0.92 & 0.98 \\
3.00 & 0.60        & 1.50 & 1.00 & 0.59 & 0.40 & 0.10 & 0.79 & 1.00 & 0.65 & 1.00 \\
3.00 & 0.60        & 2.50 & 1.00 & 0.72 & 0.56 & 0.13 & 0.71 & 1.00 & 0.59 & 1.00
\end{tabular}
\end{table}

In addition, we undertook sensitivity analysis with respect to various choices
of monitoring intervals used to generate databrick matrices. Table~\ref{tab:agglevel}
illustrates that at the relevant time-scale of interest (e.g., few seconds),
detection accuracy remains relatively unchanged.
% A moment of reflection highlights 
%that the `golden mean' between the two aggregation extremes would show the best performance;  
For optimal performance, very brief aggregation intervals for \emph{low-traffic links} are discouraged because the hash-binned arrays will be sparse.
Similarly, very large aggregation levels (e.g., minutes)  are also unwelcome
since short-duration attacks might be masked by collisions with other events.
Further, the heavy-tail modeling assumption is not suitable on large time-scales (see Section~\ref{sec:vis}).

We conclude this section be showcasing that AMON's identification may 
synergistically be coupled with the detection alerts to guide operators'
troubleshooting and mitigation efforts.
By looking at the 
database of heavy hitters reported by MJRTY BM,
network managers can readily have an IP list of candidate culprits. 
Upon detection, our algorithm outputs  a databrick bin, aimed to identify culprits. 
One can then examine the flows reported by Boyer-Moore of the sub-streams associated with the 
relevant hash bin. For example, one first sorts these 
Boyer-Moore-identified flows based on their traffic volume estimate ($P_{\rm{est}}$ in Algorithm~\ref{alg:bm}), 
and then proceeds with forensics analysis. 
In particular, for the `Library' event, in $39.7\%$
of the flagged times by Algorithm~\ref{algo:heavy-hitters},
the top ranked flow by Boyer-Moore was indeed one of the (src, dst) pairs of interest
(see Table~\ref{tab:bm_library}, top row).
For a $60.3\%$ fraction of times, the same detection method was able to identify
flow(s) of interest among the top-two reported BM flows, etc.
Similar reports are offered by the second detection algorithm at the bottom row.

\begin{table}[t]
\centering
\caption{Sensitivity on aggregation level (`Library' case)}
\vspace{-13pt}
\label{tab:agglevel}
\footnotesize
\begin{tabular}{c|cc|cc}
 & \multicolumn{2}{|c}{\begin{tabular}[c]{@{}c@{}}Fr\'echet  method\\ ($p=0.85, \lambda_\alpha=0.6$)\end{tabular}} & \multicolumn{2}{|c}{\begin{tabular}[c]{@{}c@{}}Relative Volume\\ ($\lambda_p=0.6, L=1.64)$\end{tabular}} \\
Aggregation Level & Precision & Recall & Precision & Recall \\\hline
200K NetFlow records & 0.74 & 0.52 & 0.82 & 0.53 \\
300K NetFlow records & 0.76 & 0.73 & 0.87 & 0.70 \\
400K NetFlow records & 0.92 & 0.71 & 0.81 & 0.75 \\
500K NetFlow records & 0.81 & 0.73 & 0.70 & 0.76
\end{tabular}
\vspace{-15pt}
\end{table}

\begin{table}[t!]
\centering
\caption{`Library' case study: culprit identification using  MJRTY Boyer-Moore.}
\vspace{-10pt}
\label{tab:bm_library}
\begin{tabular}{r|cccccccc}
Top-K                                 & 1      & 2    & 4    &  8    & 16   & 32   & 64  & 128 \\\hline
Fr\'echet  method (Alg.~\ref{algo:heavy-hitters}) - time fraction (\%)          & 39.7   & 60.3 & 74.0 & 90.4  & 93.2 & 98.6 & 100 & 100 \\ 
Rel. Vol. (Alg.~\ref{algo:topk-volume})  - time fraction (\%) & 42.7   & 61.3 & 74.7 & 90.7  & 93.3 & 98.7 & 100 & 100  
\end{tabular}
\vspace{-20pt}
\end{table}

\vspace{-15pt}
\subsection{Diagnosing low-volume attacks}
\label{subsec:tor}

In addition to orchestrated, volumetric attacks
that seek to overwhelm the victims with traffic, low-volume DDoS attacks can  be pernicious, albeit  
problematic to detect.  Such attacks, like the `SSDP' event previously analyzed,  rely on presumably innocuous message transmissions
to thwart standard anomaly detection methods. 
In this section, we highlight the importance of \emph{visualizations} and of
\emph{methods that detect structural patterns} (see Section~\ref{sec:detectcomm}) in traffic in revealing these low-profile nefarious actions.
As an initial example, consider Figure~\ref{fig:vis_tools} (bottom left). The bold horizontal line
in the depicted databrick is an artifact of the distributed nature of the `SSDP' attack. Operators
can easily, instantaneously  and visually observe such patterns by monitoring {AMON}'s live data products.
Further, we reiterate the important role that the connectivity algorithm played in automatically uncovering
the `SSDP' instances (see Table~\ref{tab:ssdp}).

\begin{figure*}[t!]
        \centering
        \begin{subfigure}[b]{0.32\textwidth}
                 \includegraphics[width=0.8\textwidth]{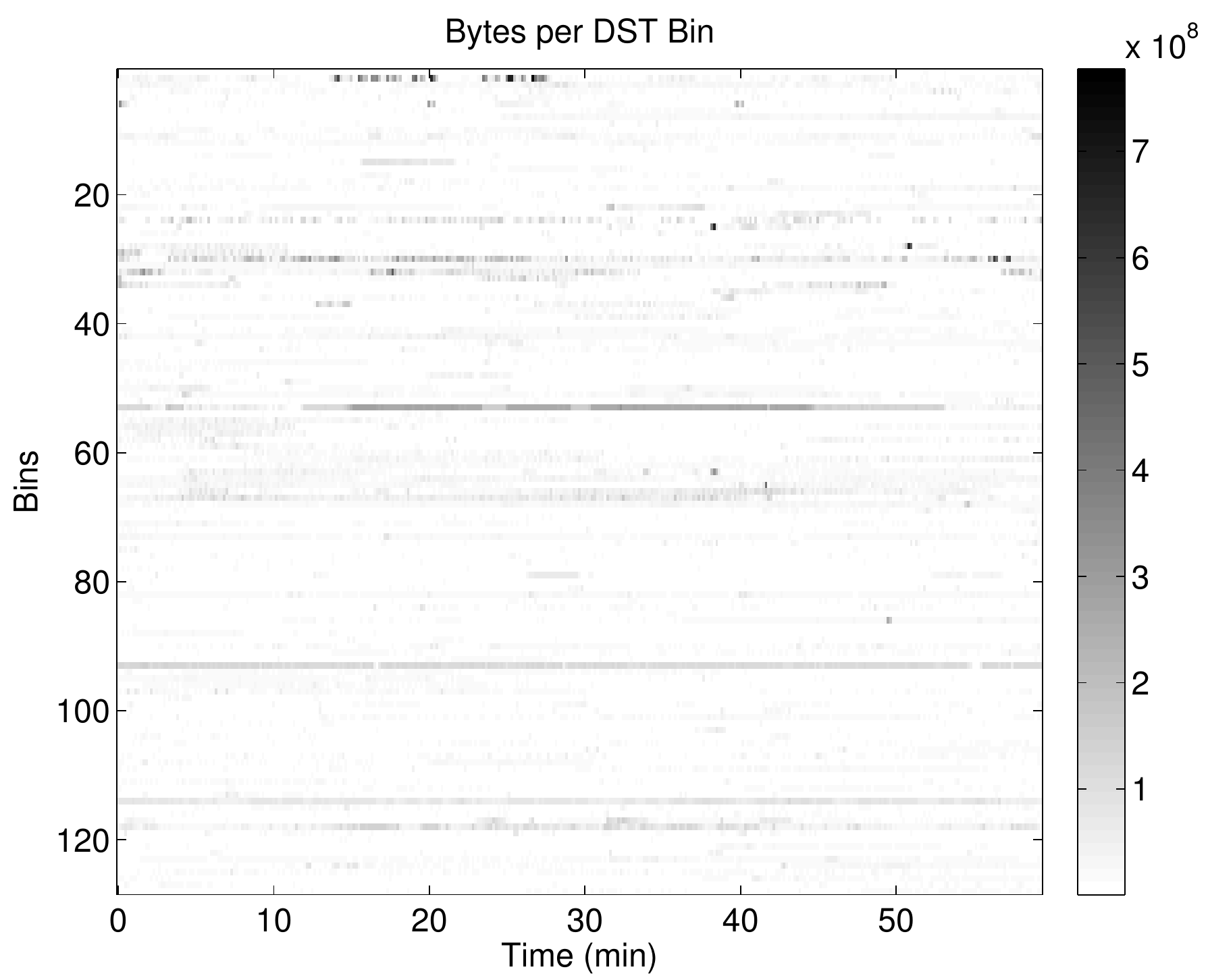}
        \end{subfigure}
        \begin{subfigure}[b]{0.32\textwidth}
               \includegraphics[width=0.8\textwidth]{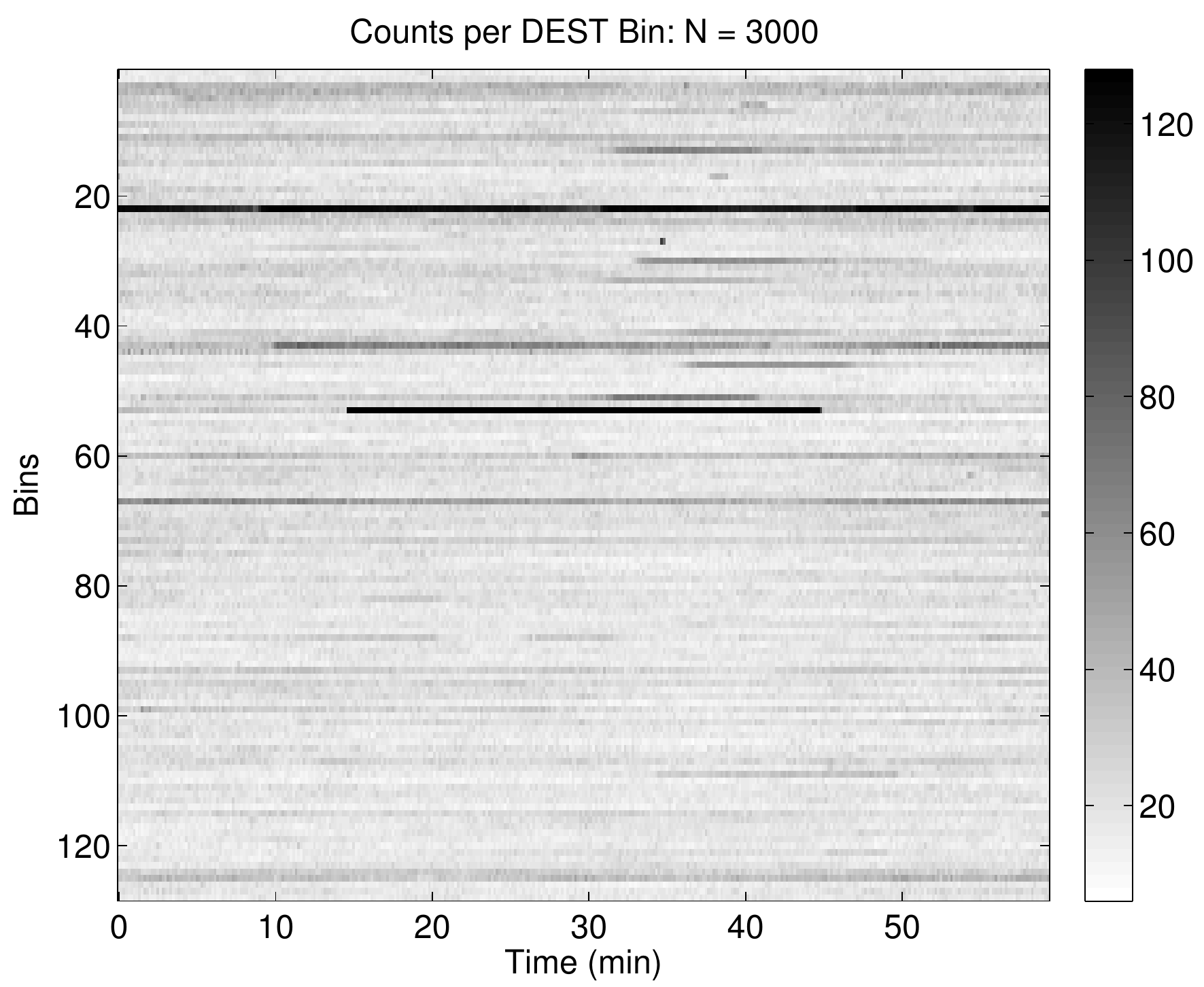}
        \end{subfigure}
        \begin{subfigure}[b]{0.32\textwidth}
                 \includegraphics[width=0.8\textwidth]{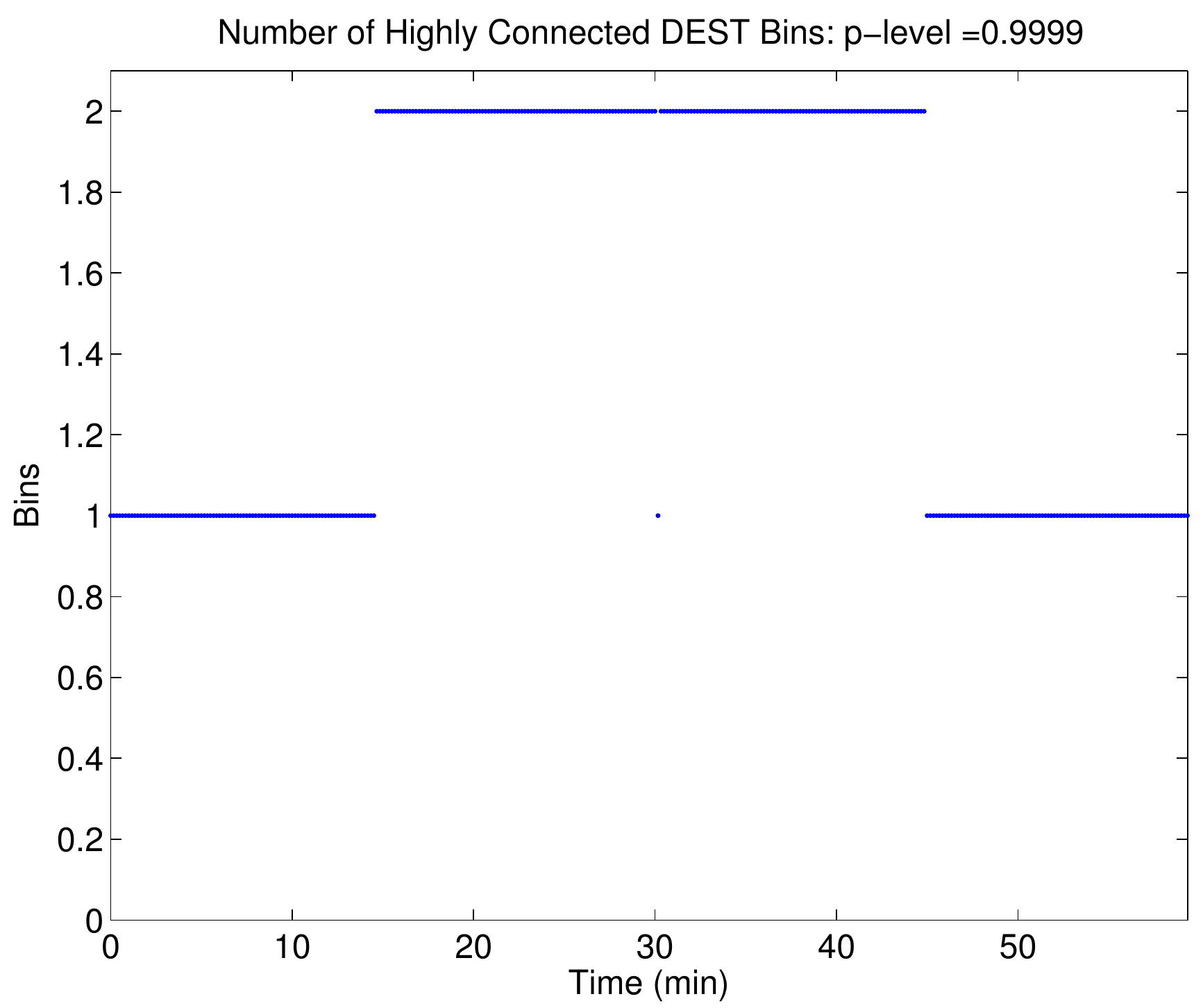}
        \end{subfigure}
 \vspace{-15pt}
 \caption{\footnotesize{Low-volume attacks (`Tor' and `SSH-scanning' events): community detection.}}
\label{fig:tor_case}
\vspace{-20pt}
\end{figure*}

\begin{figure}[!tb]
    \centering
    \begin{minipage}{.5\textwidth}
       \vspace{-40pt}
        \centering
        \hspace{-85pt}
        \begin{subfigure}[b]{0.5\textwidth}
                 \includegraphics[width=1.18\textwidth]{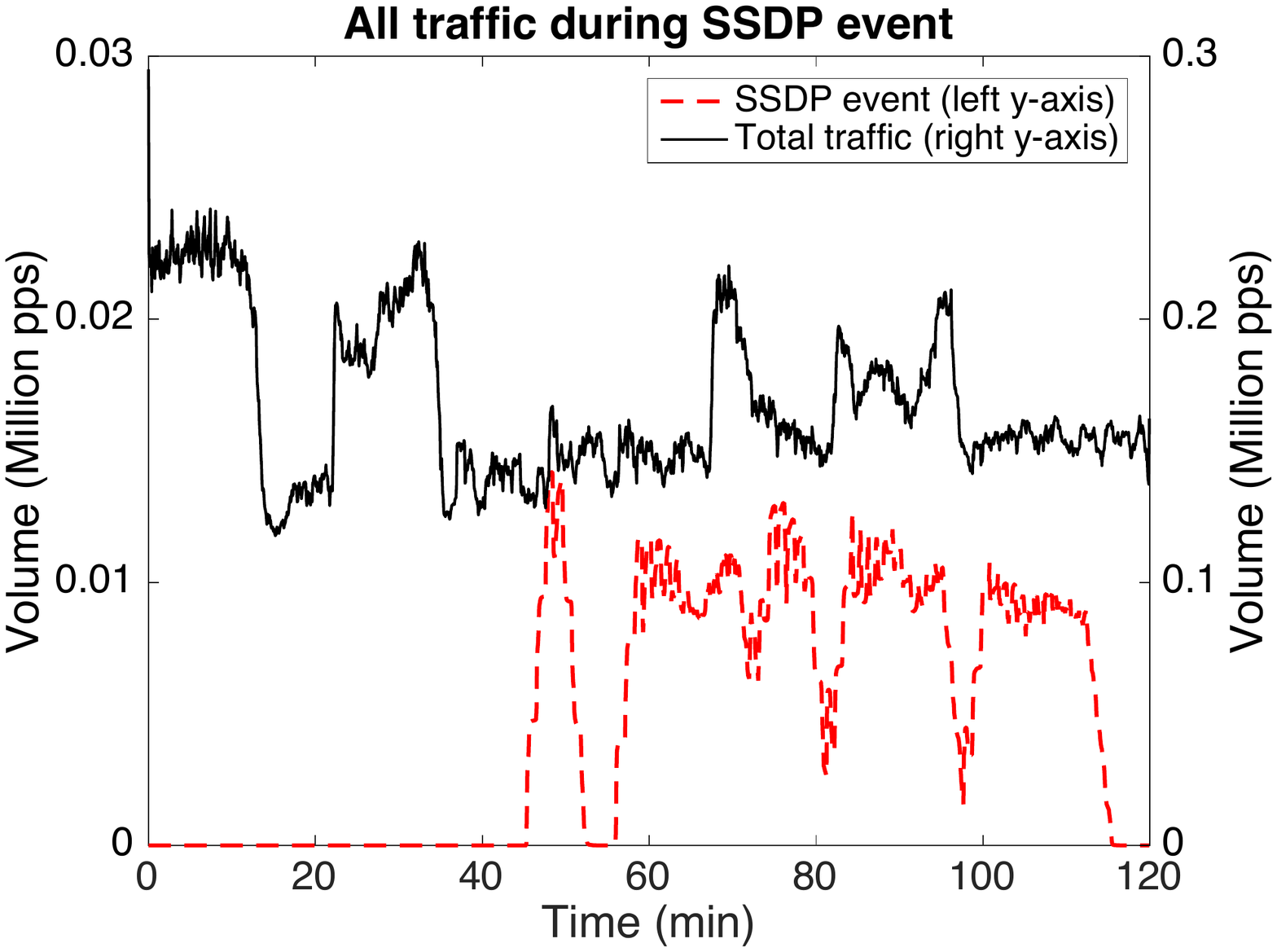}
        \end{subfigure}\\\vspace{-35pt}
        \hspace{-70pt}
        \begin{subfigure}[b]{0.5\textwidth}
               \includegraphics[width=1.05\textwidth]{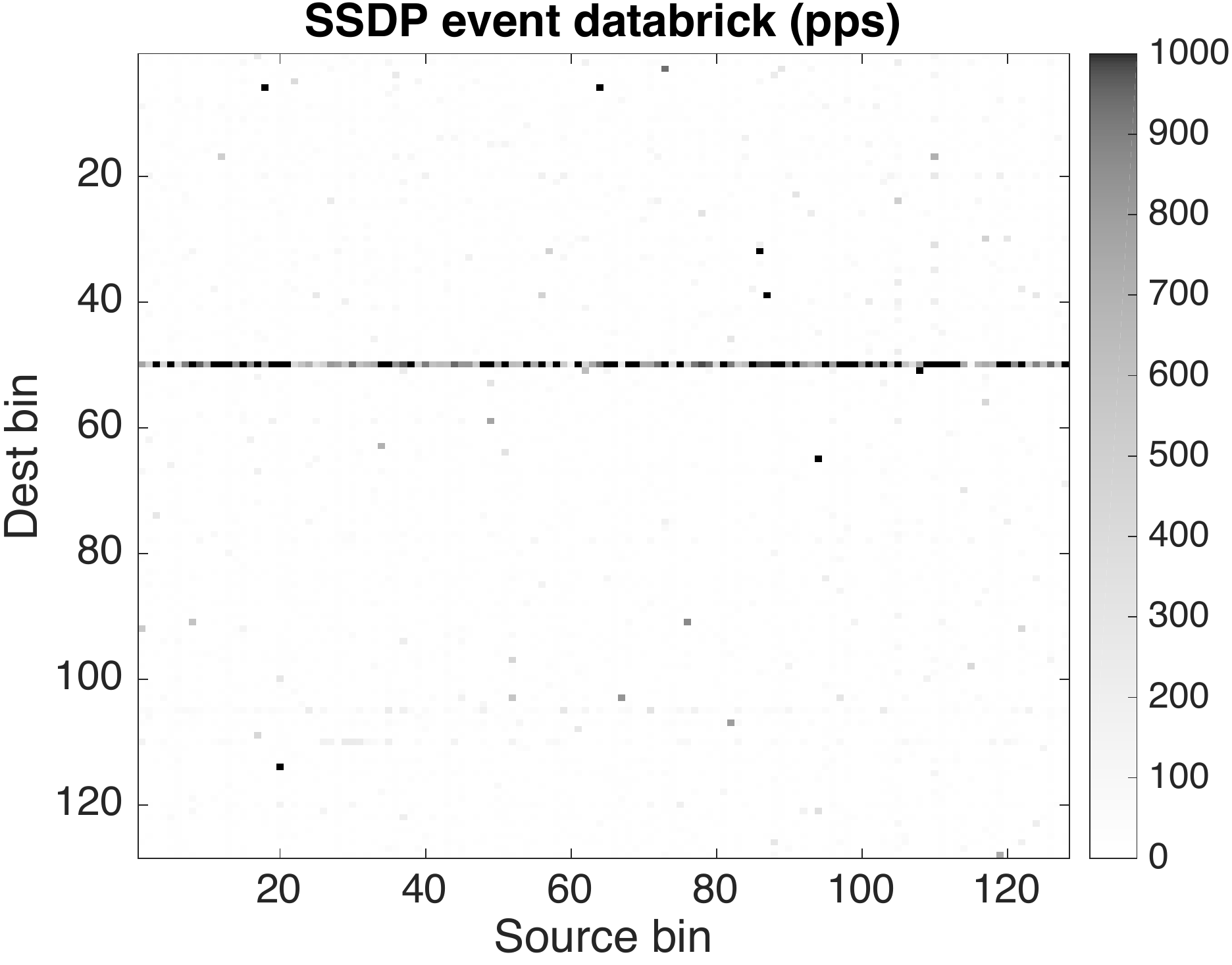}
        \end{subfigure} 
    \end{minipage}%
    \hspace{-60pt}
    \begin{minipage}{0.5\textwidth}
        \centering
      \includegraphics[width=1\textwidth]{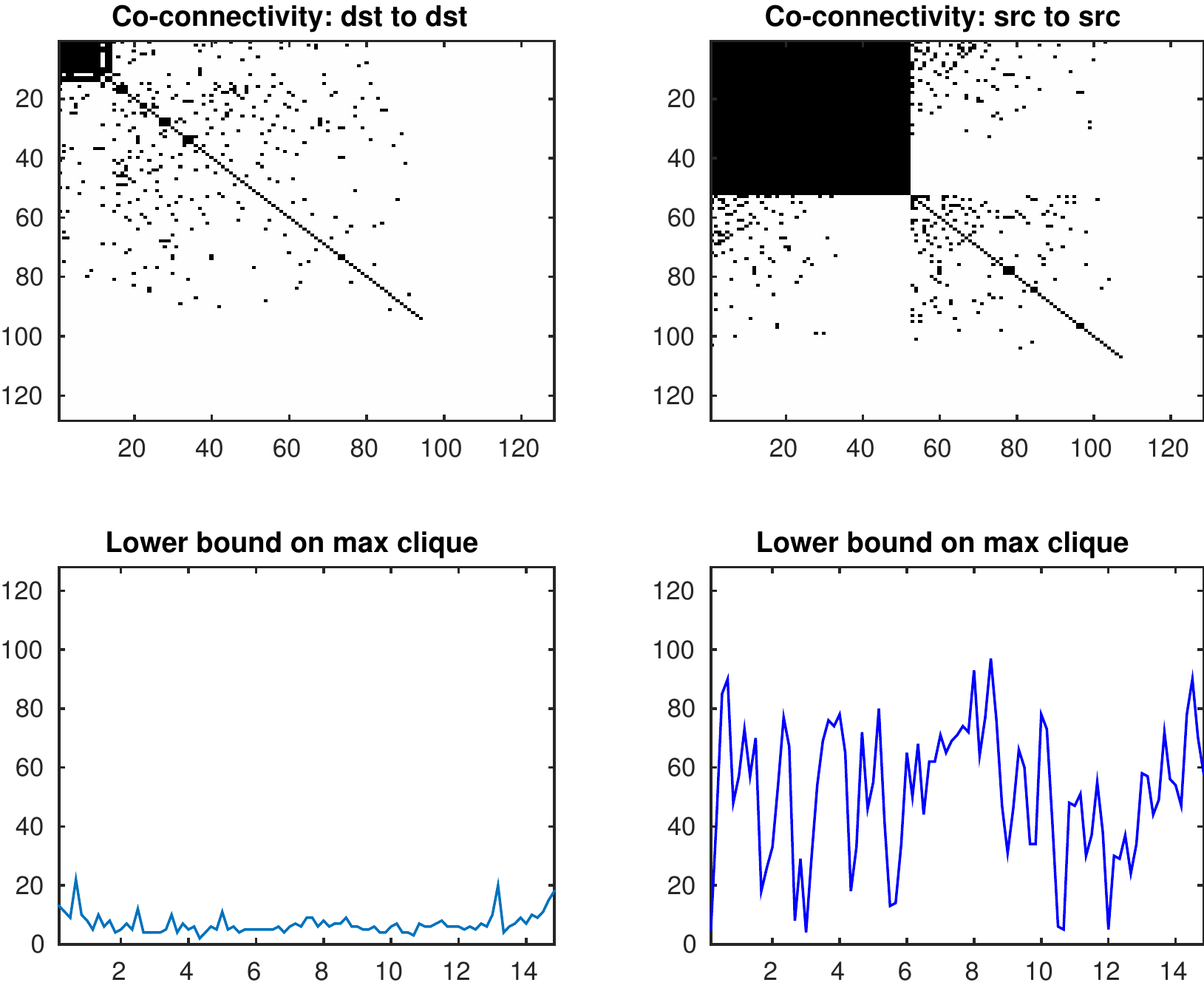}
    \end{minipage}
    \vspace{-10pt}
    \caption{\footnotesize{Visualizations readily available by our data products. \emph{Left:} Merit Network 10:00-12:00 EST, Dec 9, 2015 -- the faint `SSDP'
        event (in volume) is clearly observable in the time snapshot of the databrick matrix (horizontal line). 
         \emph{Top Right:} Adjacency matrices of co-connectivity graphs (node indices sorted by
degree---black corresponds to locations of 1's). \emph{Bottom Right:} Size of max cliques over time during
         the `Tor' case study (Section~\ref{subsec:tor}).
          By observing clique size changes in Dashboards like this, 
	 coupled with the detection method of Section~\ref{sec:detectcomm}, such 
         seemingly innocuous low-volume events are captured.}}
         \label{fig:vis_tools}
         \vspace{-30pt}
\end{figure}

Figure~\ref{fig:tor_case} depicts
another case study of this kind in which sparse traffic patterns (left panel) make 
Algorithms~\ref{algo:heavy-hitters} and~\ref{algo:topk-volume} 
to miss these events. 
As clearly seen by the in-degree counters of Figure~\ref{fig:tor_case} (middle), two
possibly suspicious events are occurring. 
Manual inspection revealed the first event to be
UDP misuse affecting
a Tor exit router within Merit, and the other (longest running)  
attempts of SSH-breaking into Michigan-located servers from IPs that belong at an autonomous system registered in 
the Asia-Pacific region.  We refer to this case study as `Tor'. %\footnote{The UDP misuse event on Tor was also reported by Merit's PeakFlow; however, the SSH brute-force attack was not.}.
The right panel illustrates that both events were flagged by our community
detection system.
This plot shows the number of highly connected destinations (i.e., high in-degree) over the
duration of an hour. The correct hash bins were also reported (22 and 53).
%Table~\ref{tab:bm_tor} shows the relevance
%of culprits reported by MJRTY BM for the reported hashed bins and time instances,
%interpreted as Table~\ref{tab:bm_library}.

Further, Figure~\ref{fig:vis_tools} (right panels) demonstrates extra visualization aids
readily available by our data products; cliques and clique sizes for the sources and destinations
\emph{co-connectivity} graphs are depicted. A co-connectivity graph for sources provides insights into
the number of {\em common destinations} between two sources; the destinations co-connectivity graph
sheds similar information for destinations.
To obtain these (undirected) graphs 
we utilize the binary matrix 
$A_t = (a_t(i,j))_{m\times m}$ (see Section~\ref{sec:detectcomm}).
The co-connectivity graph for destinations $D_t$
is efficiently obtained as $D_t := A_t\cdot A_t^T$; the
graph for sources is $S_t := A_t^T\cdot A_t$, where $A_t^T$ is the transpose of $A_t$.
Based on the co-connectivity graphs one can obtain visualizations about the cliques formed, over time.
Figure~\ref{fig:vis_tools} showcases two such snapshots. These graphs are
portrayed as their matrix adjacency representations, and we have re-arranged the node
labeling based on the node-degree in decreasing order (i.e., the first row represents the adjacency associations
of the node with the highest degree). Note the very large clique formed in src-to-src
graph. This depicts the situation in the `Tor' case study discussed above,
when a plethora of sources were contacting the same destination (the Tor exit router). 
One may also extract the maximum clique for each graph; the bottom 
row demonstrates this characteristic over time. The reader is pointed to our
supplement~\cite{kallitsis15amon_supp} for an animated version of Figure~\ref{fig:vis_tools},
where clearly one can observe the clique sizes evolving and expanding
over the duration of the  `Tor' event.

%\begin{table}[t!]
%\vspace{-10pt}
%\centering
%\caption{`Tor' case study: culprit identification using  MJRTY Boyer-Moore.}
%\vspace{-10pt}
%\label{tab:bm_tor}
%\begin{tabular}{r|cccccccc}
%Top-K                                      & 1      & 2    & 4    &  8    & 16   & 32   & 64  & 128 \\\hline
%Hash bin 22 - time fraction (\%)           & 58.7   & 86.2 & 98.0 & 100  & 100 & 100 & 100 & 100 \\ 
%Hash bin 53 - time fraction (\%)           & 82.3   & 97.8 & 99.4 & 100  & 100 & 100 & 100 & 100  
%\end{tabular}
%\vspace{-10pt}
%\end{table}

%\begin{figure}[t!]
% \centering
%  \includegraphics[width=0.5\textwidth]{./figures/tor_viz_snapshot-3.eps}
%  \vspace{-15pt}
% \caption{\footnotesize{Visualizations readily available by our data products.
%         \emph{Top:} Adjacency matrices of co-connectivity graphs (node indices sorted by
%degree -  black corresponds to locations of 1's). \emph{Bottom:} Size of max cliques over time during
%         the `Tor' case study (Section~\ref{subsec:tor}).
%          By observing clique size changes in Dashboards like this, 
%	 coupled with the detection method of Section~\ref{sec:detectcomm}, such 
%         seemingly innocuous low-volume events are captured.}}
%\label{fig:vis_tools}
%\vspace{-25pt}
%\end{figure}

\vspace{-10pt}  
\section{Conclusions}

The paper presents a novel open source monitoring architecture
suitable for multi-gigabit (i.e., 10Gbps+) network traffic streams. It is based 
on PF\_RING Zero-Copy and tailored for
deployment on commodity hardware for
troubleshooting high-impact events that may arise from malicious actions
such as DDoS attacks. It is worth noting that the NIC needed
for processing traffic data at speeds around 25Gbps (Figure~\ref{fig:pfring_chicago})
costs roughly \textbf{800 USD}, while specialized FPGA accelerated
cards or monitoring appliances cost an order of magnitude higher (above 10,000 USD).
We demonstrated the performance of our system architecture and the underlying statistical methods
on selected real-world case studies and measurements from the Merit Network. 

Our framework is extensible, and allows
for further statistical, filtering and visualization modules. Currently, we are in the process of deploying an interactive filtering
mode of operation that would enable network operators to
zoom-in and examine IP ranges of interest in real-time. As an example, consider the `Tor' and `SSH-scanning'
events of Figure~\ref{fig:tor_case} for which our methods automatically flagged bins 22 and 53. With the filtering option,
operators can rapidly zoom exclusively into the sub-stream of traffic that gets mapped into the flagged bins.
Note that due to randomization, these hash-bins are not associated with traditional IP-ranges (e.g., subnets or specific IP addresses). 
Thus, such filters cannot be implemented using existing filtering infrastructures such as BPF or hardware-based filters.

%Future developments: Software filters in zero-copy mode, Interactive monitoring via filtering,
%Automatic zoom-in on IP ranges via alarm-based filtering, 
%Geolocation of alarms: dynamic visualization and sharing of geographic location of culprits
%using ArcGIS or Google-based mapping services integrated with fusion tables.

\noindent\medskip
{\em Acknowledgments:} The authors thank L. Deri and the \texttt{ntop.org} team for providing
the necessary PF\_RING ZC licenses free-of-charge. We also thank
M. Morgan and Y. Kebede for their valuable assistance. This work is supported by 
a Department of Homeland Security Science and Technology 
Directorate FA8750-12-2-0314 grant (MK), and National Science Foundation grants
 CNS-1422078  (MK, GM) and DMS-1462368 (SS). We also thank all
 anonymous reviewers. 
 
\medskip
{\bf Supplementary material}: Software prototype, animations and multimedia provided in~\cite{kallitsis15amon_supp}.

\appendix

Proof of Proposition~\ref{p:Frechet-limit}.
This result is a simple consequence of Theorem 3.3.7, p.\ 131 in \cite{embrechts:kluppelberg:mikosch:1997}.
\begin{IEEEproof}
By the independence of the $X(i)$'s, for all fixed $x >0$, we have
$$
\P( m^{-1/\alpha} D_m(X) \le x ) = \P( X\le m^{1/\alpha} x)^m = (1- \P(X> m^{1/\alpha}x))^m. 
$$
Now, by \eqref{e:Pareto-tail} with $x$ replaced by $m^{1/\alpha}x$, we observe that 
$\P(X>m^{1/\alpha}x) \sim c/(mx^\alpha)$, as $m\to\infty$. Thus, using the fact that 
$(1- c x^{-\alpha}/m)^m \to e^{-c/x^{\alpha}},\ m\to\infty$, we conclude that
$$
 \P( m^{-1/\alpha} D_m(X) \le x ) \longrightarrow e^{-c/x^{\alpha}},\ \ \mbox{ as }m\to\infty.
$$
This implies the desired convergence in \eqref{e:p:Frechet-limit}, since 
$\P( c^{1/\alpha} Z_\alpha \le x) = e^{-c/x^\alpha},\ x>0.$
\end{IEEEproof}

Proof of Proposition~\ref{e:X-order-stat}.
\begin{IEEEproof} Part {\bf(i)} is a direct consequence of \eqref{e:X-order-stat}. 
Now, to prove {\bf (ii)}, observe that by \eqref{e:p:order-stat-i}, 
\begin{equation}\label{e:p:order-stat-1}
 \frac{V(k;m)}{V(\ell;m)} \stackrel{d}{=}  
 { \sum_{j=1}^k \overline F^{-1}(\Gamma_{j}/\Gamma_{m+1}) \over \sum_{j=1}^\ell \overline F^{-1}(\Gamma_{j}/\Gamma_{m+1}) }.
\end{equation}
By the Strong Law of Large Numbers, we have that $\Gamma_j/\Gamma_{m+1} \sim \Gamma_j/m$, as $m\to\infty$, almost surely, for all 
$j=1,\ldots,\ell$. Recall that $\ell$ is fixed. Thus, in view of \eqref{e:Pareto-tail}, $\overline F^{-1}(p) \sim (p/c)^{-1/\alpha}$, as $p\downarrow 0$, and hence for all $j=1,\ldots,\ell$, with probability one, we have
$$ 
\overline F^{-1}{\Big(}\frac{\Gamma_{j}}{\Gamma_{m+1}}{\Big)} \sim {\Big(}\frac{\Gamma_{j}}{c\Gamma_{m+1}}{\Big)}^{-1/\alpha},\ \ \mbox{ as }m\to\infty.
$$
This implies that the right-hand side of \eqref{e:p:order-stat-1} converges almost surely to
$$
 { \sum_{j=1}^k \Gamma_{j}^{-1/\alpha} (c\Gamma_{m+1})^{1/\alpha} \over \sum_{j=1}^\ell \Gamma_{j}^{-1/\alpha} (c\Gamma_{m+1})^{1/\alpha} } = W_\alpha(k,\ell),
$$
which completes the proof of \eqref{e:p:order-stat-ii}. 
\end{IEEEproof}

\bibliographystyle{IEEEtran} 
\bibliography{IEEEabrv,jsac15}
\end{document}